\documentclass[aps,prd,twocolumn,groupedaddress]{revtex4-1}
\usepackage{graphicx}
\include{amssym}

\begin{document}
\title{$1/N_c$ Nambu -- Jona-Lasinio model: Electrically charged and strange pseudoscalars.}
\author{A.\,A.\,Osipov\footnote{Email address: osipov@nu.jinr.ru}}
\affiliation{Joint Institute for Nuclear Research, Bogoliubov Laboratory of Theoretical Physics, 141980 Dubna, Russia}

\begin{abstract}
The mass formulas and decay constants of electrically charged and strange pseudoscalar mesons are analyzed within the combined framework of Nambu -- Jona-Lasinio model and the $1/N_c$ expansion up to $\mathcal O(1/N_c^2)$. The light quark masses explicitly violating $SU(3)_L\times SU(3)_R$ chiral symmetry of the strong interactions are taken to be of order $\mathcal O(1/N_c)$. The Fock-Schwinger proper-time method and the Volterra series are used to derive the effective action. A  set of sum rules is obtained that relates the phenomenological values of the masses of pseudoscalar mesons to the mass ratios of light quarks. It is shown that combining the new sum rules with the experimental data on the decay width $\eta\to 3\pi$ allows to establish limits for the ratios: $0.47<m_u/m_d<0.59$ and $18.60<m_s/m_d<19.66$. A  comparison with the results of similar calculations in $1/N_c$ chiral perturbation theory is made.
\end{abstract}

\maketitle

\section{Introduction}
Quantum chromodynamics (QCD), in the limit of a large number of colors $N_c$ \cite{Hooft:74,Witten:79}, is a successful tool to reproduce the qualitative features of strong interaction phenomena at moderate energies of the order of the $\rho$-meson mass. To succeed in the quantitative description of hadronic physics, one should rely on the effective Lagrangian approach, or on lattice calculations. Various options of the QCD inspired effective Lagrangian are discussed in the literature, differing in the content of the fields used and the rules of $N_c$ counting  \cite{Witten:80,Veneziano:80,Trahern:80,Ohta:80,Ohta:81,Leutwyler:96a,Leutwyler:96b,Taron:97,Kaiser:00,Osipov:06,Weinberg:10}. For instance, in \cite{Leutwyler:96a,Leutwyler:96b}, the properties of the lightest pseudoscalar nonet were studied by the effective Lagrangian arranged according to the powers of momenta, masses of light quarks, and $N_c$. In particular, it was assumed that the masses of light quarks are of the order $m_i=\mathcal{O}(1/N_c)$, where $i=u,d,s$ are flavors. This approach is known now as $1/N_c$ chiral perturbation theory ($1/N_c\chi$PT) \cite{Kaiser:00}  wherein the $\eta'$ meson is included consistently by means of the $1/N_c$ expansion. When $N_c\to\infty$, the axial $U(1)_A$ anomaly is absent and the pseudoscalar $SU(3)$ singlet becomes the ninth Goldstone boson associated with the spontaneous symmetry breaking of $U(3)_L \times U(3)_R \to U(3)_V$ \cite{Witten:79b,Veneziano:79}. A simultaneous chiral and $1/N_c$ expansion leads to an effective theory for the pseudoscalar nonet that is not only internally consistent but is also very useful in practice \cite{Goity:02,Bickert:20}.

In view of the fruitfulness of Leutwyler's idea to count $m_i=\mathcal{O}(1/N_c)$, we find it interesting to apply this counting rule to calculations based on the low-energy meson Lagrangian derived from the effective $U(3)_L\times U(3)_R$ symmetric four-quark interactions of the Nambu -- Jona-Lasinio (NJL) type \cite{Nambu:61a,Nambu:61b}. Our interest in the NJL model (in connection with the task of studying the properties of the pseudoscalar nonet) is due to two reasons. 

First, the model implies a specific mechanism of spontaneous chiral symmetry breaking (S$\chi$SB). Therefore, its use allows us to express a number of arbitrary parameters, known from the analysis of Leutwyler, through the  characteristics of the hadron vacuum, and thereby obtain their numerical values. This makes it possible to study in detail the four-quark mechanism of S$\chi$SB. 

Second, in obtaining the meson Lagrangian it is important to properly account for the effect of explicit violation of chiral symmetry. For this purpose, we for the first time use a new asymptotic expansion of the quark determinant \cite{Osipov:21a,Osipov:21b,Osipov:21c}, which is based on the Volterra series. This series together with the Fock-Schwinger proper-time method turns out to be an efficient tool that allows not only to isolate divergent parts of quark loop diagrams, but also to accurately reproduce the flavor structure of coupling constants of the effective meson Lagrangian. The latter circumstance is fundamental in studying the explicit violation of chiral symmetry in the NJL model. 

The method used here differs significantly from the schemes applied earlier in the NJL model to extract the consequences of explicit chiral symmetry breaking. The noncommutativity of the mass matrix of quarks with meson fields leads to an additional rearrangement of the asymptotic series in powers of proper time. As a result, the effective meson Lagrangian not only contains divergent vertices (at removal of regularization in loop integrals), but also has uniquely defined finite terms which vanish in the limit of equal masses. These terms contain, apparently, important additional information about isospin and flavor symmetry breaking, which is absent in the standard meson Lagrangian of the NJL model \cite{Volkov:84,Wadia:85, Volkov:86,Ebert:86,Bijnens:93,Osipov:17}. The study of the physical consequences of accounting for these finite contributions is a long-term goal of the approach developed here.

The pseudoscalar mesons offer an excellent ground for checking the effectiveness of the asymptotic expansion based on the Volterra representation. This concerns both the mass formulas and other low-energy characteristics of the light pseudoscalar nonet, primarily those associated with an explicit violation of chiral symmetry. The counting rule   $m_i=\mathcal{O}(1/N_c)$ makes this task more tractable for the NJL model. Indeed, any attempt to calculate the first correction to the leading-order $1/N_c$ result within the standard approach requires to account for chiral logarithms, a step that implies a major modification of the NJL model. However, if $m_i=\mathcal{O}(1/N_c)$, the contribution of chiral logarithms starts only from the order $(m_i/N_c)\ln m_i$. It is with this circumstance that the possibility of effective use of the $1/N_c$ NJL model for estimating the masses and other characteristics of the pseudoscalar nonet of mesons is connected. And it is in this case that the Volterra series plays the main role in describing the effects of explicit chiral symmetry breaking.

In this article, we deal with electrically charged and strange pseudoscalars. The neutral states $\pi^0$, $\eta$, and $\eta'$ are considered in a separate article. This is due both to the volume of the material presented and to the convenience of its perception.

The article is organized as follows. In Sec.\,\ref{s2}, we briefly describe the method for deriving the effective meson Lagrangian based on four-quark interactions, and demonstrate how the Volterra representation is embedded in the general scheme of the Fock-Schwinger proper-time method. In Sec.\,\ref{s3}, we discuss modifications related with the $1/N_c$ treatment of the NJL gap equation. The mixing between pseudoscalar and axialvector fields is considered in Sec.\,\ref{s4}. The kinetic terms of the meson effective Lagrangian are considered in Sec.\,\ref{s5}. Here we obtain the decay constants of pseudoscalars by rescaling the corresponding collective variables. The masses of the charged pion and strange pseudoscalars are discussed in Sec.\,\ref{s6}. The Gasser-Leutwyler ellipse and other sum rules relating pseudoscalar masses with light quark masses are discussed in Sec.\,\ref{s7}. Our numerical estimates and discussing the dependence of results on regularization used are given in Sec.\,\ref{s8}. We summarize in Sec.\,\ref{Conclusions}.

\section{Effective Lagrangian}
\label{s2}
Four-quark interactions are widely used to describe the mechanism of S$\chi$SB and construct the effective action of mesons at moderate energies \cite{Volkov:84,Volkov:86,Ebert:86,Osipov:17,Bijnens:93}
\begin{equation}
\label{Lagr1}  
\mathcal L=\bar q(i\gamma^\mu\partial_\mu -m)q+\mathcal{L}_{4q}.
\end{equation}
Hereinafter in the text we use the standard Lorentz metric convention $g_{\mu\nu}=\mbox{diag}(1,-1,-1,-1)$, where the indices $\mu$, $\nu$ take values in the set $\{0,1,2,3\}$; $\gamma^\mu$ are the Dirac gamma-matrices, $\bar q=(\bar u,\bar d,\bar s)$ is a flavor triplet of Dirac 4-spinors with $\bar u=u^\dagger \gamma^0$, and $m$ is a diagonal matrix $m=\mbox{diag}(m_u,m_d,m_s)$ containing the masses of current up, down and strange quarks. The Lagrange density describing four-quark interactions has the form $\mathcal{L}_{4q}=\mathcal L^{(0)}+\mathcal L^{(1)}$, where the sum consists of $G=U(3)_L\times U(3)_R$ chiral symmetric four-quark operators with spin zero and one respectively
\begin{eqnarray}
 \mathcal L^{(0)} &=& \frac{G_S}{2} \left[ (\bar q\lambda_a q)^2+(\bar q i\gamma_5\lambda_a q)^2  \right], \\
 \mathcal L^{(1)} &=&- \frac{G_V}{2} \left[ (\bar q\gamma^\mu\lambda_a q)^2+(\bar q \gamma^\mu\gamma_5\lambda_a q)^2  \right], 
\end{eqnarray}
where $a=0,1,\ldots, 8$, the matrix $\lambda_0=\sqrt{2/3}$ and $\lambda_i$ are the eight Gell-Mann matrices of $SU(3)$. The coupling constants $G_S$ and $G_V$ have dimensions (mass)$^{-2}$ and can be fixed from the meson mass spectrum. 

The spin-0 short-range attractive force between light quarks $\sim G_S(\bar q\lambda_a q)^2$ is responsible for the $S\chi SB$. If this interaction is sufficiently strong $G_S\geq G_{\mbox{\footnotesize{cr}}}$, it can rearrange the vacuum, and the ground state becomes superconducting, with a nonzero quark condensate. As a result, nearly massless current quarks become effectively massive constituent quarks. The short-range interaction can then bind these constituent quarks into mesons.

For a theory described by the Lagrangian density (\ref{Lagr1}), the vacuum to vacuum amplitude is given by functional integration 
\begin{eqnarray}
\label{Zq}
Z&\!=\!\!&\int [dq][d\bar q] \exp \left( i\!\!\int\!\! d^4x \,\mathcal L\right)  \\
\label{Zsq}
  &\!=\!\!&\int [dq][d\bar q] [ds_a][dp_a][dv_a^{\mu}][da_a^{\mu}]\exp \left( i\!\!\int\!\! d^4x \,\mathcal L' \right)\!, \nonumber
\end{eqnarray}
where 
\begin{eqnarray}
\label{Lagr2}
\mathcal L' &=& \bar q [i\gamma^\mu \partial_\mu +s+i\gamma_5 p 
                        +\gamma^\mu (v_\mu +\gamma_5 a_\mu )] q \nonumber \\
&-& \frac{\mbox{tr}[(s+m)^2+p^2]}{4G_S} +\frac{\mbox{tr}(v_\mu^2+a_\mu^2)}{4G_V}. 
\end{eqnarray}
The new Lagrangian $\mathcal L'$ has the same dynamical content as $\mathcal L$ since if we perform a functional integration over collective variables $s_a$, $p_a$, $v^\mu_a$ and $a^\mu_a$ in (\ref{Zsq}) we obtain the original expression (\ref{Zq}).  Notice that $s=s_a\lambda_a$, $p=p_a\lambda_a$, $v^\mu=v^\mu_a\lambda_a$ and $a^\mu=a^\mu_a\lambda_a$. 

In the world of zero quark bare masses $m=0$, $\mathcal L'$ is invariant under global $G=U(3)_L\times U(3)_R$ transformations. In particular, the group $G$ acts on the quark fields as follows 
\begin{equation}
q'=(P_RV_R+P_LV_L)q=e^{i\alpha}e^{i\gamma_5\beta} q,
\end{equation}
where the projection operators $P_{R,L}$ are $P_R=(1+\gamma_5)/2$, $P_L=(1-\gamma_5)/2$. It is convenient to choose finite unitary transformations $V_{R,L}\in G$ in the form of a product of two exponents $V_R=e^{i\alpha}e^{i\beta}$, $V_L=e^{i\alpha}e^{-i\beta}$, where $\alpha=\alpha_a\lambda_a$, $\beta=\beta_a\lambda_a$ and the parameters $\alpha_a$ and $\beta_a$ are real. Then it follows that 
\begin{eqnarray}
s'+ip'&=&V_L(s+ip)V^\dagger_R, \nonumber \\
v'_\mu+a_\mu' &=& V_R(v_\mu+a_\mu)V_R^\dagger, \nonumber \\
v'_\mu -a_\mu' &=& V_L (v_\mu -a_\mu)V_L^\dagger. 
\end{eqnarray} 

Let us use the freedom of choice of dynamical variables in (\ref{Zsq}) to carry out the transition to a nonlinear realization of chiral symmetry for Goldstone particles. To do this, we represent the complex  $3\times 3$ matrix $s+ip$ as the product of the unitary matrix $\xi$  and the Hermitian matrix $\tilde\sigma$ 
\begin{equation}
s+ip=\xi\tilde\sigma\xi.
\end{equation}
From the covariance of this expression under the action of the group $G$, it follows that the matrices $\xi$ and $\tilde\sigma$ are transformed as
\begin{equation}
\xi'=V_L\xi h^\dagger = h\xi V_R^\dagger, \quad \tilde\sigma'=h\tilde\sigma h^\dagger,
\end{equation}
where $h$ is a unitary compensating transformation belonging to the maximal subgroup $H\subset G$, leaving the vacuum invariant, and arising under the action of the chiral group $G$ on the coset representative $\xi$ of the $G/H$ manifold. In these variables we have $\bar q(s+i\gamma_5 p)q=\bar Q\tilde\sigma Q,$ where the new quark fields are given by $Q=(\xi P_R+\xi^\dagger P_L)q$. A nonlinear realization of $G$ becomes a linear representation when restricted to the subgroup $H$ \cite{Wess:69a,Wess:69b}. Indeed, one can see that the field $Q$ transforms as the fundamental representation of the subgroup $H$: $Q'=hQ$.  

Having done similar redefinitions in the rest of the Lagrangian (\ref{Lagr2}), we find $ \mathcal L' \to \mathcal L''$, where
\begin{eqnarray}
\label{ELM}  
  \mathcal L''&=&\bar Q (i\gamma^\mu d_\mu -M+\sigma ) Q +\frac{1}{4G_V} \mbox{tr}(V_\mu^2+A_\mu^2) \nonumber \\
  &-&\frac{1}{4G_S} \mbox{tr}\left[\sigma^2-\{\sigma ,M\} +(\sigma-M)\Sigma \right].
\end{eqnarray}  
Notice the replacement $\tilde\sigma=\sigma -M$ made in (\ref{ELM}). The matrix $M$ is diagonal $M=\mbox{diag} (M_u,M_d,M_s)$, and its elements are considered as the masses of constituent quarks $Q$. We assume on this step that chiral symmetry is realized in the Nambu-Goldstone sense (as $m_i\to 0$), i.e., heavy constituent masses result from dynamic symmetry breaking and are controlled by the gap equation (see below Eq.\,(\ref{gapEq})). In turn, $\sigma (x)$ describes quantum fluctuations of the $\tilde\sigma$ field around a physical vacuum. 

The corresponding collective variables for vector, axial-vector, scalar and pseudoscalar fields are given by Hermitian matrices $V_\mu=V_\mu^a\lambda_a$, $A_\mu=A_\mu^a\lambda_a$, $\sigma=\sigma_a\lambda_a$, $\phi=\phi_a\lambda_a$, where 
\begin{eqnarray}
V_\mu &=& \frac{1}{2} \left[\xi (v_\mu +a_\mu )\xi^\dagger +\xi^\dagger (v_\mu -a_\mu )\xi \right], \nonumber \\
A_\mu &=& \frac{1}{2} \left[\xi (v_\mu +a_\mu )\xi^\dagger -\xi^\dagger (v_\mu -a_\mu )\xi \right], \nonumber \\
\Sigma &=&\xi m \xi+\xi^\dagger m \xi^\dagger , \quad  \xi=\exp \left(\frac{i}{2}\,\phi\right).
 \end{eqnarray}
The pseudoscalar field $\phi$ is dimensionless, later on, when passing to the field functions of physical states, it will acquire the required dimension of mass. The vector $V_\mu$ and axial-vector $A_\mu$ fields are chosen to transform as $V_\mu'=hV_\mu h^\dagger$, $A_\mu'=hA_\mu h^\dagger$. 

The symbol $d_\mu =\partial_\mu -i\Gamma_\mu$ in (\ref{ELM}) denotes the covariant derivative, where
\begin{eqnarray}
\label{gammamu}
  &&\Gamma_\mu = \xi^{(+)}_\mu +V_\mu+\gamma_5\left( \xi_\mu^{(-)}+A_\mu  \right), \\
  && \xi_\mu^{(\pm )}= \frac{i}{2} \left( \xi\partial_\mu \xi^\dagger \pm \xi^\dagger \partial_\mu \xi  \right).
  \end{eqnarray}  
It is easy to establish that $\Gamma_\mu$ is a connection on $G/H$ satisfying the standard transformation rules under the local action of $G$
\begin{equation}
\Gamma_\mu' = h\Gamma_\mu h^\dagger +ih\partial_\mu h^\dagger.
\end{equation} 
To be precise, this is how $\xi_\mu^{(+)}$ is transformed, the other fields in $\Gamma_\mu$  are the covariant objects with similarity transformations.    
  
To exclude quark degrees of freedom in the functional integral 
\begin{equation}
\label{Z3}
Z\!=\!\!\int [dQ][d\bar Q] [d\sigma_a] \mathcal D\mu [\phi_a] [dV_a^{\mu}][dA_a^{\mu}] e^{\, i\!\!\int\!\! d^4x \,\mathcal L''}\!,
\end{equation}
it is necessary to integrate over the quark variables. Before we do this, let us clarify that the $G$ invariant measure $\mathcal D\mu [\phi_a]$ in $Z$ is related to the curvature of the $G/H$ group manifold parametrized by the pseudoscalar variables $\phi_a$. It is easy to find an explicit expression for this differential form, but we will not need it in what follows. Therefore, we better proceed directly to the calculation of the real part of the effective meson Lagrangian taking the integral over quark fields. The result is a functional determinant 
\begin{equation}
\label{logdet1}  
  W_E =\ln |\det D_E| 
         = -\int\limits^\infty_0\!\frac{dt}{2t}\,\rho_{t,\Lambda}\,\mbox{Tr}\left(e^{-t D_E^\dagger D_E^{}}\right),
\end{equation}
representing a real part of the one-loop effective action in Euclidian (E) space as the integral over the proper-time $t$. Here, $D=i\gamma^\mu d_\mu -M+\sigma \to D_E$. Note that the rules we use to continue to Euclidean space are standard and can be found, for instance, in \cite{Osipov:21c}. The symbol "Tr" denotes the trace over Dirac $(D)$ $\gamma$-matrices, color $(c)$ $SU(3)$ matrices, and flavor $(f)$ matrices, as well as integration over coordinates of the Euclidean space: $\mbox{Tr}\equiv \mbox{tr}_I \int\! d^4x_E$, where $I=(c,D,f)$. The trace in the color space is trivial: It leads to the overall factor $N_c=3$. 

The dependence on matter fields in $D_E$ after switching to the Hermitian operator
\begin{equation}
D_E^\dagger D_E^{}=M^2 -d^2+Y
\end{equation}
is collected in the $3\times 3$ matrix $Y$ 
\begin{eqnarray}
Y&=&\sigma^2-\{\sigma,M\} +i[\gamma_\alpha (\sigma -M), d_\alpha ]  \nonumber \\
&+&\frac{1}{4}[\gamma_\alpha, \gamma_\beta ] [d_\alpha, d_\beta ],
\end{eqnarray}  
and the covariant derivative $d_\alpha$
\begin{eqnarray}
d_\alpha &=&\partial_\alpha +i\Gamma_\alpha, \\
\Gamma_\alpha &=&V_\alpha -\xi_\alpha^{(+)} +\gamma_{5E} (A_\alpha -\xi_\alpha^{(-)} ),
\end{eqnarray}
where $\Gamma_\alpha$ is a connection in a curved factor space of Goldstone fields (in four-dimensional Euclidean space a convention that the Greek indices $\alpha$ and $\beta$ run from $1$ to $4$ is used). 

In (\ref{logdet1}), to reguralize quadratic and logarithmic divergences in the proper-time integrals we introduce the kernel $\rho_{t,\Lambda}$ which provides two subtractions on the same mass scale $\Lambda$
\begin{equation}
\rho_{t,\Lambda}=1-(1+t\Lambda^2)e^{-t\Lambda^2},
\end{equation}
which in the NJL model was used, for instance, in \cite{Osipov:85}. The ultraviolet cutoff $\Lambda$ characterizes the scale of $S\chi SB$, i.e., above this scale four-quark interactions disappear and QCD becomes perturbative. Obviously, the value of $\Lambda$ depends on the regularization scheme used, and generally varies in the interval  $0.65-1.3\,\mbox{GeV}$ \cite{Osipov:85,Klevansky:92}. In the present paper, we apply the proper-time regularization with $\Lambda =1.1\,\mbox{GeV}$. This value, as will be shown above, is phenomenologically justified and is consistent with an estimate of the chiral symmetry breaking scale $\Lambda_{\chi SB}\leq 4\pi f_\pi$ \cite{Georgi:84}, where $f_\pi=92.2\,\mbox{MeV}$ is the pion decay constant.  

The functional trace in (\ref{logdet1}) can be evaluated by the Schwinger technique of a fictitious Hilbert space. The use of a plane wave with Euclidian 4-momenta $k$, $\langle x|k\rangle $, as a basis greatly simplifies the calculations (details, for instance, are given in \cite{Osipov:21c}) and leads to the representation of the functional trace by the integrals over coordinates and 4-momenta 
\begin{equation}
\label{logdet-2}  
     W_E=\! - \!\!\int\!\!\frac{d^4x d^4k}{(2\pi )^4}\, e^{-k^2}\!
            \!\! \int\limits^\infty_0\!\!\frac{dt}{2t^3}\,\rho_{t,\Lambda}\,
          \mbox{tr}_I\! \left(e^{-t(M^2+A)}\right)\! .
\end{equation}
The self-adjoint operator $A$ is given by 
\begin{equation}
\label{A}
 A= -d^2 -2ik d / \sqrt{t} +Y,
\end{equation} 
where a summation over four-vector indices is implicit.  
 
To advance further in our calculation of (\ref{logdet-2}), we use the Volterra series 
\begin{equation}
\label{alg}
e^{-t(M^2+A)}=e^{-tM^2}\!\left[1+\sum_{n=1}^\infty (-1)^n f_n(t,A) \right],
\end{equation}
where the expression in the square brackets is the time-ordered exponential $\mbox{OE}[-A](t)$ of $A(s)= e^{sM^2}\! A\, e^{-sM^2}$, accordingly 
\begin{equation}
f_n(t,A)=\!\int\limits_0^t\!\! ds_1\!\!\int\limits_0^{s_1}\!\! ds_2 \ldots \!\!\!\!\int\limits_0^{s_{n-1}}\!\!\!\! ds_n A(s_1) A(s_2) \ldots A(s_n).
\end{equation} 
This series generalizes the standard large mass expansion of the heat kernel to the case of unequal masses. If masses are equal, this formula yields the well-known large mass expansion with standard Seeley-DeWitt coefficients $a_n(x,y)$ \cite{Ball:89}. In fact, formula (\ref{alg}) is an extension of the Schwinger's method used to isolate the divergent aspects of a calculation in integrals with respect to the proper time \cite{Schwinger:51,DeWitt:65} to the non-commutative algebra $[M,A]\neq 0$ (see also \cite{Feynman:51}). 

Inserting Eq.\,(\ref{alg}) into (\ref{logdet-2}) with the following integrations over four-momenta $k_\alpha$ and the proper-time $t$ one finds -- after the continuation to Minkowski space -- the one-quark-loop (1QL) contribution to the effective meson Lagrangian in the form of the asymptotic series 
\begin{equation}
\label{WE}
\mathcal L_{\mbox{\scriptsize 1QL}} = - \frac{N_c}{32\pi^2} \sum_{n=1}^{\infty}\,\mbox{tr}\,b_n(x,x),
\end{equation}
where coefficients $b_n(x, x)$ depend on the meson fields and quark masses. These coefficients contain the full information about both the effective meson vertices and the corresponding coupling constants. The first two coefficients are \cite{Osipov:21c}
\begin{eqnarray}
\label{cb1}
\mbox{tr}\,b_1&=&\mbox{tr}_{Df} \left[-J_0\circ Y -\frac{1}{4} (\Delta J_0\circ \Gamma_\mu )\Gamma^\mu  \right], \\
\label{cb2}
\mbox{tr}\,b_2&=&\mbox{tr}_{Df} \left[\frac{Y}{2} J\!\circ Y-\frac{1}{12}\Gamma^{\mu\nu} (J\circ \Gamma_{\mu\nu})\right] \nonumber\\
&+&\mbox{tr}_D\,\Delta b_2,
\end{eqnarray}
where $\Gamma^{\mu\nu}=\partial^\mu\Gamma^\nu -\partial^\nu\Gamma^\mu - i[\Gamma^\mu , \Gamma^\nu ]$.

For convenience, along with the usual matrix multiplication, we use here the non-standard Hadamard product \cite{Styan:73}, which is the matrix of elementwise products $(A\circ B)_{ij} =A_{ij} B_{ij}$. The Hadamard product is commutative unlike regular matrix multiplication, but the distributive and associative properties are retained. In addition, the following notations are used. 

The proper-time integral $J_0$ is considered as a diagonal matrix with elements given by $(J_0 )_{ij} = \delta_{ij} J_0 (M_i)$, where 
\begin{equation}
\label{J0}
J_0(M_i)\! =\!\!\int\limits_0^\infty \!\frac{dt}{t^2}\,\rho_{t,\Lambda}\, e^{-tM_i^2}\!=\!
\Lambda^2-M^2_i\ln\left(1+\frac{\Lambda^2}{M^2_i}\right).  
\end{equation}
This matrix collects contributions of one-loop Feynman diagrams, known as a "tadpole".

The other set of proper-time integrals in (\ref{cb2}) is given by the matrix $J$ with elements $J_{ij}=J_1(M_i,M_j)$ 
\begin{eqnarray}
\label{J-ij}
&&J_{ij}=\frac{1}{\Delta_{ij}}\left[J_0(M_j)-J_0(M_i)\right]  \\
&&=\frac{1}{\Delta_{ij}}\left[M_i^2 \ln \left(1+\frac{\Lambda^2}{M_i^2}\right)-M_j^2 \ln \left(1+\frac{\Lambda^2}{M_j^2}\right)\right] \nonumber
\end{eqnarray}
with $\Delta_{ij}=M_i^2-M_j^2$. If the masses are equal $M_i=M_j$ it gives the diagonal elements
\begin{equation}
J_{ii}\equiv J_1(M_i) =\ln\left(1+\frac{\Lambda^2}{M^2_i}\right)-\frac{\Lambda^2}{\Lambda^2+M^2_i}.
\end{equation}
It can be seen from this expression that the integral diverges logarithmically at $\Lambda\to\infty$. To stress this, we use the subscript $1$ in labelling of such integrals to distinguish them from the quadratic divergence of integrals $J_0$.

The last set of proper-time integrals which we will need in the following is given by the matrix $\Delta J_0$ with elements $(\Delta J_0)_{ij}=\Delta J_0(M_i,M_j)$. Here 
\begin{equation}
\label{DJ0ij}
\Delta J_0(M_i,M_j)=2J_0(M_i,M_j)-J_0(M_i)-J_0(M_j)
\end{equation}
 and
\begin{eqnarray}
\label{J0ij}
&&\!\!\!\!\!\!\! J_0(M_i,M_j)=\frac{\Lambda^2}{2} +\frac{\Lambda^4}{2\Delta_{ij}} \ln\frac{\Lambda^2+M_i^2}{\Lambda^2+M_j^2} \nonumber \\
&&\!\!\!\!\!\!\!\!\! -\frac{1}{2\Delta_{ij}}\left[M_i^4\ln\left(1+\frac{\Lambda^2}{M_i^2}\right)-M_j^4\ln\left(1+\frac{\Lambda^2}{M_j^2}\right)\right]\!.
\end{eqnarray}
In the coincidence limit $M_i\to M_j$, we have $$\lim_{M_i\to M_j} J_0(M_i,M_j)=J_0(M_i)\,,$$ and therefore $(\Delta J_0)_{ii}=0$. In the case of unequal masses, the difference (\ref{DJ0ij}) is finite (at $\Lambda\to\infty$) and thus gives us an example of a contribution that does not occur in the standard approach to the NJL model.

Recall that the standard meson NJL Lagrangian absorbs only divergent parts of one-loop quark diagrams \cite{Volkov:86,Ebert:86,Kikkawa:76}. They are represented by the first term in (\ref{cb1}) and the first two terms in (\ref{cb2}). Contrary to the standard approach, the coefficients $b_1$ and $b_2$ additionally have many (about hundred) finite contributions of Feynman diagrams compactly assembled in the second term in (\ref{cb1}) and third term in (\ref{cb2}). Each of them vanishes in the chiral limit; therefore they break either isotopic or $SU(3)_f$ symmetry. The appearance of these new vertices is understandable - they arise as a finite difference when subtracting two or more divergent integrals with different masses. Both the structure and the coupling constant of any finite vertex are uniquely fixed by the Volterra series. Since for the tasks considered here, we do not need the expression of $\mbox{tr}_D\,\Delta b_2$, we do not give its explicit form, but refer the interested reader to the work \cite{Osipov:21c}, where the corresponding expressions were obtained. 

It is easy to understand why there are no finite terms in the standard approach. The reason is contained in the treatment of the heat kernel $\exp[-t(M^2+A)]$. To find the asymptotics of this object, one usually separates a commutative matrix $\mu$: $M^2+A=\mu^2+A+(M^2-\mu^2)$ to factorize it from the exponent. As a result, the expansion of the heat kernel contains only standard Seeley-DeWitt coefficients
\begin{equation}
\label{alg8}
e^{-t(M^2+A)}=e^{-t\mu^2}\sum_{n=0}^\infty t^n a_n(x,x).
\end{equation}
The mass scale $\mu$ is arbitrary. For example, this parameter may be identified with the average constituent quark mass $\mu =\mbox{tr} M/3$ \cite{Ebert:86} or with the constituent quark mass in the chiral limit $\mu =M_0$ \cite{Bijnens:93}. In both cases, the integration over proper time $t$ in (\ref{alg8}) yields couplings $\propto J_n(\mu )$ which are not sensitive to the flavor content of quark-loop integrals. The explicit  violation of chiral symmetry is carried out only due to the corresponding part of $Y$ and the term $M^2-\mu^2$. This approach leads to the different pattern of the flavor symmetry breaking and does not contain the finite terms as well because usually only the first two field-dependent terms in (\ref{alg8}) ($n=1,2$) are considered. 

An attempt made in \cite{Ebert:86} to restore the flavor dependence of the coupling constants by replacing the divergent integrals $J_n(\mu )$ by the expressions following from the direct calculations of corresponding Feynman graphs is inconsistent mathematically, although being correct qualitatively. In such a way it is impossible to trace the pattern of an explicit chiral symmetry breaking without distorting it. The Volterra series (\ref{alg}) not only gives a rigorous foundation of the substitutions made in \cite{Ebert:86}, but also associates with them a definite finite part.   

So, as a result of the calculations performed, we finally arrive to the effective meson Lagrangian given by 
\begin{eqnarray}
\label{EL}
\mathcal L''&\to& \mathcal L_{\mbox{\scriptsize eff}}=\mathcal L_{\mbox{\scriptsize 1QL}} +\frac{1}{4G_V}\mbox{tr}_f(V_\mu^2+A_\mu^2) \nonumber \\
&-&\frac{1}{4G_S} \mbox{tr}_f\! \left[\sigma^2-\{\sigma, M\}+(\sigma -M)\Sigma\right].
\end{eqnarray}
This Lagrangian contains all the information about chiral symmetry breaking, including effects induced by unequal quark masses. In what follows we will be interested only in the part of this Lagrangian that is responsible for the physics of pseudoscalar mesons.

\section{Gap equation and $1/N_c$ expansion}
\label{s3}
First let us exclude from the effective Lagrangian (\ref{EL}) the linear in $\sigma$ term (the tadpole). The corresponding contributions are contained in $\mathcal L_{\mbox{\scriptsize 1QL}}$ and the last term in (\ref{EL}). Singling them out, e.g.,
\begin{equation}
\mbox{tr}_{Df}\, (-J_0\circ Y)\to 8\!\!\sum_{i=u,d,s}\!\! J_0(M_i) M_i \sigma_i , 
\end{equation}
we arrive at the Lagrangian  
\begin{equation}
\label{tadpoleL}
\mathcal L_\sigma =\sum_{i=u,d,s}\!\! \sigma_i\left[\frac{M_i-m_i}{2G_S}-\frac{N_c}{4\pi^2} M_i J_0(M_i)  \right].
\end{equation} 

Requiring that the tadpole term vanishes, we obtain a self-consistency equation
\begin{equation}
\label{gapEq}
M_i \left(1-\frac{N_c G_S}{2\pi^2} J_0(M_i)\right)=m_i,
\end{equation}
which relates the mass of light quark $m_i$ to the mass of heavy constituent quark $M_i$. This equation can be rewritten in terms of the quark condensate
\begin{equation}
\label{QC}
\langle \bar q\lambda_i q\rangle =-\frac{M_i-m_i}{2G_S}.
\end{equation}

In the strong coupling regime
\begin{equation}
\label{SCR}
G_S\Lambda^2>\frac{2\pi^2}{N_c}=6.58, 
\end{equation}
each of three $(i=u,d,s)$ equations (\ref{gapEq}) has a nontrivial solution, which describes a gap in the spectrum of fermions. This solution signals that the ground state becomes superconducting, with a nonzero quark condensate. Knowing that spontaneous breaking of chiral symmetry is present in QCD at large $N_c$ \cite{Hooft:74,Witten:79}, we conclude that $G_SN_c=\mbox{const}$, and $\Lambda\sim {\mathcal O}(1)$ in the large-$N_c$ limit.

Let us emphasize the difference between the approaches associated with the two alternative assumptions made for the current quark mass counting rule at large $N_c$. 

If we assume that $m_i=\mathcal O(1)$, then both parts of the gap equation (\ref{gapEq}) are present in leading order in $1/N_c$. As a consequence, one should look for an exact solution of the gap equation. The result can be also presented as a series in powers of current quark masses \cite{Osipov:92}. In this case, it is always possible to estimate the accuracy of the expansion used by comparing the truncated result to the exact solution.  

The counting rule $m_i=\mathcal O(1/N_c)$, yields that the right hand side of Eq.\,(\ref{gapEq}) at leading $1/N_c$-order (LO) tends to zero; chiral symmetry is restored $m_i = 0$; the masses of all constituent quarks are equal to the same value $M_0$, which is determined by the equation 
\begin{equation}
\label{gap2}  
  1-\frac{N_cG_S}{2\pi^2}J_0(M_0)=0.
\end{equation}  

The nontrivial solution of the gap equation $M_i(m_i)$ must be understood now as an asymptotic series 
\begin{equation}
\label{AS}
M(m)=\sum_{k=0}^{n}M_k(m)+{\mathit o}(M_n(m)), 
\end{equation}
where at $m\to 0$, i.e., at $N_c\to\infty$, each next term in the right-hand side of (\ref{AS}) is an infinitesimally small in comparison with the previous one. The essential difference of the series (\ref{AS}) from the standard solution of the gap equation is that $M_k(m)$ may collect additional contributions from the meson loop diagrams. For instance, $M_0$ may obtain a $1/N_c$ correction from a scalar tadpole graph which contributes to $M_1(m)$. In the following, for the sake of simplicity, we shall restrict our consideration to the mean-field approximation, i.e., neglect the quantum effects due to scalar fields. On the contrary, a pseudoscalar tadpole gives a leading (in the chiral limit $m_i\to 0$) non-analytic contribution only at higher order $m/N_c \ln m$ to the term $M_2(m)$, and therefore does not affect the first two terms of the asymptotic series (\ref{AS}). Correspondingly, at the next, $M_3(m)$, step, it is necessary to take into account the contribution of two-loop meson diagrams, and so on. This implies a corresponding modification of both the effective potential and the gap equation. Thus, even in the mean-field approximation a full account of current quark masses by the naive summation of the Taylor series does not work here, because it does not account for essential contributions of chiral logarithms arising at higher powers of light quark masses. 

For specific calculations, it is necessary to limit oneself only to those terms that do not exceed the accuracy of the calculations performed. So, up to the next to leading order (NLO) correction included, we can write
\begin{equation}
\label{expm}  
M_i(m_i)=M_0+M'(0) \, m_i +\mathit o (M_2(m)).
\end{equation}  
Here we suppose that the quark condensate is the leading order parameter of the spontaneously broken symmetry. It follows then from (\ref{QC}) that $M_0=\mathcal O(N_c^0)$. As we will see later, the $1/N_c$-correction to the mass formulae of charged pseudoscalars is small, which speaks in favor of the hypothesis just adopted.   

The second term of the asymptotic expansion (\ref{expm}) can be easily determined by differentiating Eq.\,(\ref{gapEq}) under the assumption that $m_i$ are independent variables. At the first step, we have $M_1(m)=M'(0)m$, where
 \begin{equation}
\label{Mprime}
M'(0)=\frac{\pi^2}{N_cG_S M_0^2 J_1(M_0)}\equiv a.
\end{equation}
Since $J_1(M_0)$ is a monotonically decreasing positive definite function of $M_0$ in the region $M_0>0$, we conclude that $a>0$. It follows that the first correction increases the mass of the constituent quark.

\section{PA-mixing}
\label{s4}
To address the physical pseudoscalar fields, it is necessary to eliminate the mixing of pseudoscalars with axial-vector fields (PA-mixing), and also to separate the kinetic part of the free Lagrangian of pseudoscalars in $\mathcal L_{\mbox{\scriptsize 1QL}}$. 

The first goal is achieved by redefining the axial vector field \cite{Morais:17}
\begin{equation}
\label{PA}
A_\mu=A_\mu'-\kappa_A\circ \xi_\mu^{(-)},
\end{equation}
where the nonet of axial-vector fields is given by 
$$
A_\mu = 
\left(\begin{array}{ccc}
f_{u\mu} &\sqrt 2 a_{1\mu}^+ & \sqrt 2 K^+_{1A\mu} \\
\sqrt 2 a_{1\mu}^- & f_{d\mu} & \sqrt 2 K^0_{1A\mu} \\
\sqrt 2 K^-_{1A\mu} & \sqrt 2 \bar K^0_{1A\mu} &  f_{s\mu} 
\end{array}\right).
$$
The nine pseudoscalar fields are collected in a hermitian matrix 
$$
\phi = \phi_a \lambda_a =
\left(\begin{array}{ccc}
\phi_u &\sqrt 2 \pi^+ & \sqrt 2 K^+ \\
\sqrt 2 \pi^- & \phi_d & \sqrt 2 K^0 \\
\sqrt 2 K^- & \sqrt 2 \bar K^0 &  \phi_s 
\end{array}\right),
$$
where the diagonal elements are
\begin{eqnarray}
\label{uds-038}
\phi_u&=&\phi_3+\frac{1}{\sqrt 3}\left(\phi_8 +\sqrt 2\phi_0 \right), \nonumber \\
\phi_d&=&-\phi_3+\frac{1}{\sqrt 3}\left(\phi_8 +\sqrt 2\phi_0 \right), \nonumber \\
\phi_s&=& \frac{1}{\sqrt 3}\left(\sqrt 2\phi_0 -2\phi_8 \right).
\end{eqnarray}

After the replacement (\ref{PA}) PA-mixing terms contained in $\mathcal L_{\mbox{\scriptsize 1QL}} $ and in the second term of (\ref{EL}) can be canceled by an appropriate choice of a matrix $\kappa_A$. To demonstrate this, it is necessary to consider the following terms of the effective meson Lagrangian  
\begin{eqnarray}
\label{pa1}
\mathcal L_{\mbox{\scriptsize 1QL}}^{(b_1)}&\to& \frac{N_c}{32\pi^2}\left\{ (\Delta J_0)_{ud} \left[(1-\kappa_{Aud})\partial_\mu \pi^+ +2a_{1\mu}^{'+}\right] \right. \nonumber \\
&\times & \left[(1-\kappa_{Aud})\partial_\mu \pi^- +2a_{1\mu}^{'-}\right] \nonumber \\
&+&(\Delta J_0)_{us} \left[(1-\kappa_{Aus})\partial_\mu K^+ +2K_{1A\mu}^{'+}\right]  \nonumber \\
&\times & \left[(1-\kappa_{Aus})\partial_\mu K^- +2K_{1A\mu}^{'-}\right] \nonumber \\
&+&(\Delta J_0)_{ds} \left[(1-\kappa_{Ads})\partial_\mu K^0 +2K_{1A\mu}^{'0}\right]  \nonumber \\
&\times &\left. \left[(1-\kappa_{Ads})\partial_\mu \bar K^0 +2\bar K_{1A\mu}^{'0}\right] \right\},
\end{eqnarray}
where the symbol $(b_1)$ indicates that the considered contribution is due to the coefficient $b_1$. 

The next contribution owes its origin to the coefficient $b_2$, namely its part described by the first term of Eq.(\ref{cb2})
\begin{eqnarray}
\label{pa2}
&&\mathcal L_{\mbox{\scriptsize 1QL}}^{(b_2)}\to \frac{N_c}{16\pi^2}\left\{ (M_u+M_d)^2 J_1(M_u,M_d)  \right. \nonumber \\
&&\times \left[(1-\kappa_{Aud})\partial_\mu \pi^+ +2a_{1\mu}^{'+}\right] \left[(1-\kappa_{Aud})\partial_\mu \pi^- +2a_{1\mu}^{'-}\right] \nonumber \\
&&+(M_u+M_s)^2 J_1(M_u,M_s)  \left[(1-\kappa_{Aus})\partial_\mu K^+ +2K_{1A\mu}^{'+}\right]  \nonumber\\
 &&\times    \left[(1-\kappa_{Aus})\partial_\mu K^- +2K_{1A\mu}^{'-}\right] \nonumber \\
&&+(M_d+M_s)^2 J_1(M_d,M_s)  \left[(1-\kappa_{Ads})\partial_\mu K^0 +2K_{1A\mu}^{'0}\right]  \nonumber \\
&&\times \left. \left[(1-\kappa_{Ads})\partial_\mu \bar K^0 +2\bar K_{1A\mu}^{'0}\right] \right\}.
\end{eqnarray}

It remains to take into account the last contribution related to the PA-mixing, which arises due to the second term in (\ref{EL}). This contribution is 
\begin{eqnarray}
\label{pa3}
&&-\frac{1}{2G_V} \left(\kappa_{Aud} a_{1\mu}^{'-} \partial_\mu \pi^+      
                              +\kappa_{Aus} K_{1A\mu}^{'-} \partial_\mu K^+    \right. \nonumber \\ 
&& \ \ \ \ \ \ \ \ \ \ \  \left. +\kappa_{Ads} \bar K_{1A\mu}^{'0} \partial_\mu K^0 \right) +h.c.  
\end{eqnarray}

Collecting the results (\ref{pa1}), (\ref{pa2}) and (\ref{pa3}), we find that the matrix $\kappa_A$ is symmetric $(\kappa_A)_{ij}=(\kappa_A)_{ji}$, with elements given by
\begin{equation}
\label{kA-1}
\kappa_{Aij}^{-1}=1+\frac{8\pi^2}{N_cG_V[2(M_i+M_j)^2 J_{ij}+\Delta J_{0ij}]}.
\end{equation}
In the chiral limit this result coincides with the standard NJL approach, but it differs in the general case. It follows then, that $G_V=\mathcal O(1/N_c)$, and, in particular, for Eq.\,(\ref{Mprime}) we find 
\begin{equation}
a=\frac{G_V}{G_S}\left(\kappa^{-1}_{A0}-1\right),
\end{equation}
where the index $0$ means that the function of the quark masses $(\kappa^{-1}_{A})_{ij}$ is calculated in the chiral limit $m_i\to 0$, i.e., 
\begin{equation}
\kappa^{-1}_{A0}=1+\frac{\pi^2}{N_cG_VM_0^2J_1(M_0)}=\frac{Z_0}{Z_0-1}.
\end{equation}
The last equality relates $\kappa^{-1}_{A0}$ with the constant $Z$  \cite{Osipov:85b,Volkov:86,Ebert:86} commonly used in the NJL model to get rid of the PA-mixing effect, $Z_0=\lim_{m_i\to 0} Z$.

The first two terms in the expansion of Eq.\,(\ref{kA-1}) in powers of $1/N_c$ are given by 
\begin{equation}
\label{kAij-1exp}
\kappa_{Aij}^{-1}=\kappa^{-1}_{A0}\left[1-\frac{m_i+m_j}{2M_0}\left(a-\delta_M\right) \right]+\mathcal O(1/N_c^2),
\end{equation}
where
\begin{equation}
\label{dM}  
\delta_M=a\left\{1-2(1-\kappa_{A0})\left[1- \frac{\Lambda^4 J_1(M_0)^{-1}}{(\Lambda^2+M_0^2)^2}  \right]\right\}.
\end{equation}  
Notice that $\Delta J_{0ij}$ contributes to (\ref{kAij-1exp}) only starting from the $1/N_c^2$ order. As we will show soon, $\delta_M$ determines the first order correction to the current algebra result for masses of electrically charged and strange pseudoscalars. It is a function of four parameters $\Lambda$, $G_S$, $G_V$ and $M_0$, which determine the structure of the hadron vacuum.

\section{Kinetic terms and decay constants}
\label{s5}
Our next task is to obtain the kinetic part of the free Lagrangian of pseudoscalar fields. To do this, we need the already known expressions (\ref{pa1}), (\ref{pa2}) and, in addition, one should write out the corresponding contribution of the second term in (\ref{EL}), that was omitted in (\ref{pa3}) 
\begin{eqnarray}
&&\frac{1}{4G_V} \left(\kappa_{Aud}^2 \partial_\mu\pi^+\partial_\mu\pi^- 
                                 + \kappa_{Aus}^2 \partial_\mu K^+\partial_\mu K^- \right. \nonumber \\
&&\ \ \ \ \ \  \left.  + \kappa_{Ads}^2 \partial_\mu \bar K^0\partial_\mu K^0 \right).
\end{eqnarray}

Collecting all these contributions, one finds, for instance in the case of charged pions, that the kinetic term is given by
\begin{eqnarray}
\mathcal L_{\mbox{\scriptsize kin}}^{\pi^+\pi^-}&=&   \partial_\mu\pi^+\partial_\mu\pi^-  
\left\{  \frac{\kappa^2_{Aud}}{4G_V} + \frac{N_c}{32\pi^2} (1-\kappa_{Aud})^2\right.   \nonumber \\
 &\times&\left. [2(M_u+M_d)^2J_1(M_u,M_d)+\Delta J_{0ud} ]\frac{}{}\right\} \nonumber \\
&=& \left( \frac{\kappa_{Aud}}{4G_V} \right)\,  \partial_\mu\pi^+\partial_\mu\pi^- .
\end{eqnarray}

To give this expression a standard form, one should introduce the physical pion fields $\pi^\pm_{\mbox{\tiny ph}}$
\begin{equation}
\label{fpi}
\pi^{\pm} =\sqrt{\frac{4G_V}{\kappa_{Aud}}}\, \pi^{\pm} _{\mbox{\tiny ph}}=\frac{1}{f_\pi} \, \pi^{\pm} _{\mbox{\tiny ph}}.
\end{equation}
The dimensional parameter $f_\pi$ is nothing else than the weak decay constant of a charged pion. Similar calculations in the case of kaons give the values of the corresponding weak decay constants
\begin{equation}
\label{fK}
f_{K^\pm}= \sqrt{\frac{\kappa_{Aus}}{4G_V}}, \quad f_{K^0}= \sqrt{\frac{\kappa_{Ads}}{4G_V}}. 
\end{equation} 

The resulting expressions require a more detailed discussion. 

First, they differ from the standard result of the NJL model, where the constant $f_\pi$ is estimated through the quark analog of the Goldberger-Treiman relation. The latter is a result of current algebra. Therefore, it is valid only in the leading order of expansion in current quark masses. It can be easily shown that in the chiral limit the formula (\ref{fpi}) coincides with the result of the standard approach. 

Second, using Eq.\,(\ref{kAij-1exp}), one obtains from (\ref{fpi}) and (\ref{fK}) the first order corrections to the current algebra result 
\begin{equation}
\label{fij}
f_{ij}=F\left(1+\frac{m_i+m_j}{4M_0}(a-\delta_M )\right),
\end{equation}
where
\begin{equation}
F=\sqrt{\frac{\kappa_{A0}}{4G_V}}=\mathcal O(\sqrt{N_c})
\end{equation}
is the pion weak decay constant in the chiral limit. In particular, it follows then that
\begin{equation}
\label{K/pi}
\frac{f_{K^\pm}}{f_\pi}=1+\frac{m_s-m_d}{4M_0}(a-\delta_M ).
\end{equation}

It is instructive to compare our result (\ref{fij}) with the result of $1/N_c\chi$PT. In this approach the corrections to $f_\pi$ and $f_K$ are determined by the constant $K_6=4B_0 L_5^r/F^2$ \cite{Gasser:85}, where the low-energy coupling constant $L_5^r$ counts of $\mathcal O(N_c)$, and the constant $B_0=\mathcal O(N_c^0)$ is related to the quark condensate. Such comparison yields 
\begin{equation}
\label{K6}
\frac{a-\delta_M}{4M_0}\leftrightarrow  K_6.
\end{equation}  
That demonstrates the full agreement between the two approaches at this stage.

It is well known that the numerical values of the ratio (\ref{K/pi}) calculated by the various groups of authors using NJL model lie between $1.02$ and $1.08$ \cite{Klevansky:92} and thus underestimate the experimental value $f_K/f_\pi =1.19$. As we show below, formula (\ref{K/pi}) perfectly reproduces the experimental value, as it also takes place in $1/N_c\chi$PT.

\section{Mass formulas and current quark masses}
\label{s6}
Let's establish now the mass formulas of $\pi^\pm$, $K^\pm$, $K^0$ and $\bar K^0$ mesons. To do this, we need the corresponding contribution arising from the last term of the Lagrangian (\ref{EL})
\begin{eqnarray}
\label{psm}
\frac{1}{4G_S}\mbox{tr}_f\, M\Sigma \to &-&\frac{1}{4G_S}\left[ (M_u+M_d)(m_u+m_d)\pi^+\pi^- \right. \nonumber \\
&+&(M_u+M_s)(m_u+m_s)K^+K^- \nonumber \\
&+&\left. (M_d+M_s)(m_d+m_s)\bar K^0K^0  \right].
\end{eqnarray}
Note that Lagrangian $\mathcal L_{\mbox{\scriptsize 1QL}}$ does not contribute to the pseudoscalar masses. 

Now, after redefinitions of fields in Eq.\,(\ref{psm}), we finally arrive to the result  
\begin{eqnarray}
\mathcal L_{\mbox{\scriptsize mass}}&=&-\frac{G_V}{G_S}
\left[ \frac{1}{\kappa_{Aud}}(M_u+M_d)(m_u+m_d)\pi_{\mbox{\tiny ph}}^+\pi_{\mbox{\tiny ph}}^- \right. \nonumber \\
&+&\frac{1}{\kappa_{Aus}}(M_u+M_s)(m_u+m_s)K_{\mbox{\tiny ph}}^+K_{\mbox{\tiny ph}}^- \nonumber \\
&+&\left.\frac{1}{\kappa_{Ads}} (M_d+M_s)(m_d+m_s)\bar K_{\mbox{\tiny ph}}^0K_{\mbox{\tiny ph}}^0  \right].
\end{eqnarray}
It follows then that the masses are 
\begin{eqnarray}
  \label{pi}
  &&\bar M_{\pi^\pm}^2=\frac{ 1}{4G_S f^2_{ud}} (M_u+M_d)(m_u+m_d), \\
  \label{K+}
  &&\bar M_{K^\pm}^2=\frac{1}{4G_S f^2_{us}} (M_u+M_s)(m_u+m_s), \\
  \label{K0}
  &&\bar M_{K^0}^2=\frac{1}{4G_S f^2_{ds}} (M_d+M_s)(m_d+m_s). 
\end{eqnarray}  
Here and below, the overline indicates that the masses were obtained without taking into account electromagnetic corrections. It should be emphasized that Eqs.\,(\ref{pi})-(\ref{K0}) differ from similar expressions obtained in \cite{Volkov:86,Ebert:86} and in other available works where the NJL model has been used. In our result the sum of the current quark masses is factorized, i.e., the Gell-Mann--Oakes--Renner relation \cite{Oakes:68} is already satisfied at this level. Obviously, all above NJL-based results coincide in the limit of exact $SU(3)_f$ symmetry. However, when calculating the first $1/N_c$ correction, these approaches lead to different results. In favor of the Eqs.\,(\ref{pi})-(\ref{K0}), as we will now see, is their agreement with the results of similar calculations made in the $1/N_c\chi$PT. 

Expanding expressions (\ref{pi})-(\ref{K0}) in $1/N_c$ series, one can not only obtain the known result of the current algebra \cite{Weinberg:77} 
\begin{eqnarray}
\label{GOR}
\bar\mu^2_{\pi^\pm}&=&B_0 (m_u+m_d), \nonumber \\
\bar\mu^2_{K^\pm}&=&B_0 (m_u+m_s), \nonumber \\
\bar\mu^2_{K^0}&=&B_0 (m_d+m_s), 
\end{eqnarray}
where the constant $B_0$ is related to the quark condensate $\langle\bar qq\rangle_0\equiv\langle\bar uu\rangle_0=\langle\bar dd\rangle_0=\langle\bar ss\rangle_0$
\begin{equation}
B_0=\frac{2G_VM_0}{G_S\kappa_{A0}}=\frac{M_0}{2G_SF^2}=-\frac{\langle\bar qq\rangle_0}{F^2},
\end{equation}
but also to move further and obtain the first order correction  
\begin{eqnarray}
\label{pi1}
  &&\bar m_{\pi^+}^2=\bar\mu^2_{\pi^+}\left(1+\frac{m_u+m_d}{2M_0}\, \delta_M \right), \\
\label{K+1}  
  &&\bar  m_{K^+}^2=\bar\mu^2_{K^+}\left(1+\frac{m_u+m_s}{2M_0}  \, \delta_M \right), \\
 \label{K01} 
  &&\bar  m_{K^0}^2=\bar\mu^2_{K^0} \left(1+\frac{m_d+m_s}{2M_0} \, \delta_M \right).
\end{eqnarray}  
These relations agree with those of $1/N_c\chi$PT \cite{Leutwyler:96a}, i.e., the following correspondence between parameters takes place  
\begin{equation}
\label{dmch}
\frac{\delta_M}{2M_0}\leftrightarrow K_3=8 \frac{B_0}{F^2}\left(2L_8^r-L_5^r\right).
\end{equation}

As pointed out in \cite{Kaplan:86}, in chiral perturbation theory $L_8^r$ cannot be determined on purely phenomenological grounds. Treating $L_8^r$ as a free parameter, one may obtain both the positive and negative sign for the difference $2L_8^r-L_5^r$. In the framework of the $1/N_c\chi$PT, Leutwyler managed to establish the generous low bound for the range where a truncated $1/N_c$ expansion leads to meaningful results: $2L^r_8-L_5^r>0$. Based on formulas (\ref{K6}) and (\ref{dmch}), it is easy to express these low-energy constants in terms of the NJL model parameters
\begin{equation}
\label{L8}
L_5^r=\frac{(a-\delta_M)G_SF^4}{8M_0^2}, \quad L_8^r=\frac{aG_SF^4}{16M_0^2}.
\end{equation}

As a consequence of (\ref{dmch}), we find the $1/N_c$ correction $\Delta_M$ considered in \cite{Leutwyler:96a}  
\begin{equation}
\label{DeM}
\Delta_M = \frac{8}{F^2}(M_K^2 - M_\pi^2)\left(2L_8^r-L_5^r\right) \leftrightarrow \frac{m_s-\hat m}{2M_0}\,\delta_M,
\end{equation}
where $\hat m=(m_u+m_d)/2$. The value of $\Delta_M$ characterizes the degree of breaking of $SU(3)_f$ symmetry. Although $\Delta_M$ cannot be calculated within the framework of $1/N_c\chi$PT, the estimate $0 < \Delta_M \leq 0.13$ was obtained in \cite{Leutwyler:97} based on additional reasonable considerations. 

\begin{figure}
\includegraphics[width=0.45\textwidth]{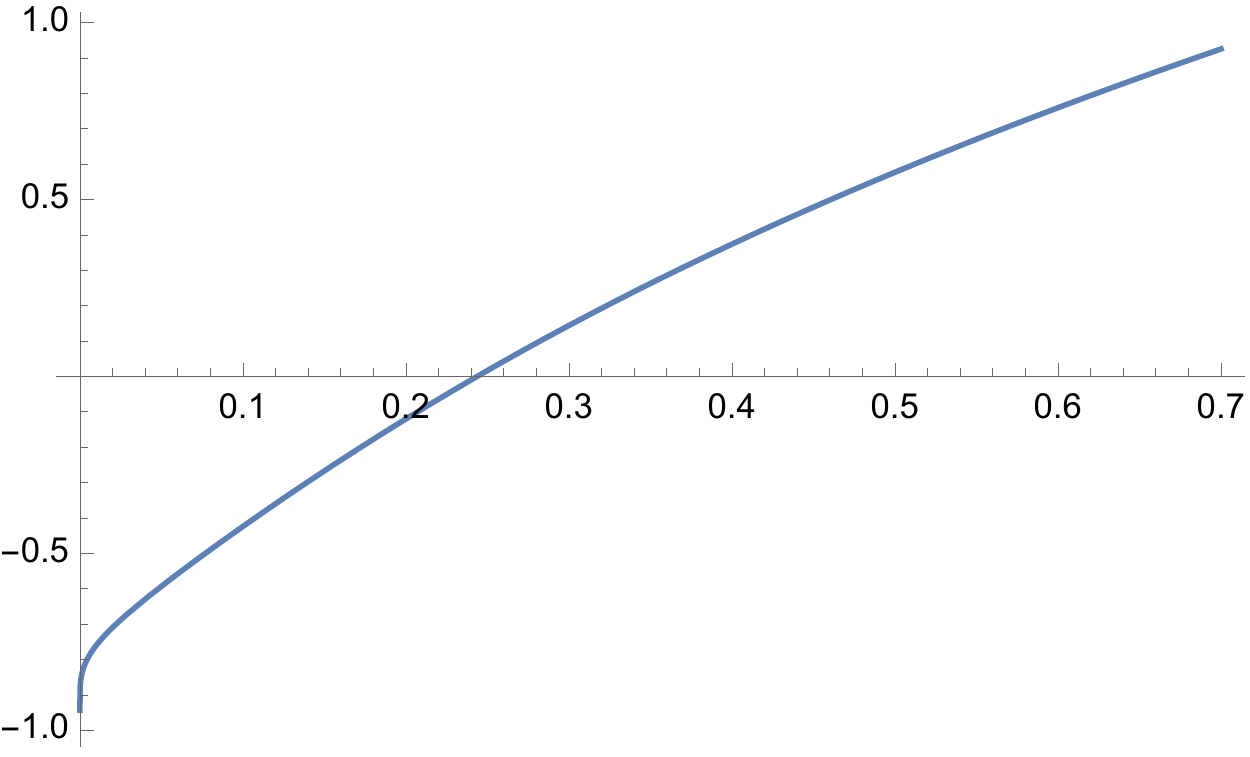}
\caption{The ratio $\delta_M/a$ (see Eq.\,(\ref{dM})) is shown as a function of $M_0$ (in GeV) at fixed values of $\Lambda = 1.1\,\mbox{GeV}$, $G_S = 6.4\,\mbox{GeV}^{-2}$, and $G_V = 3.6\,\mbox{GeV}^{-2}$. For such parameter setting, the point $M_{0\mbox{\tiny min}}=0.244\,\mbox{GeV}$ satisfies both the requirement $\delta_M=0$, and Eq.\,(\ref{gap2}).}
\label{fig1}      
\end{figure}

In the model studied here, the sign of $\delta_M$ coincides with the sign of the ratio $\delta_M/a$ in Eq.\,(\ref{dM}), which, as it is shown in Fig.\ref{fig1}, is a monotonically increasing function of the variable $M_0$ (at $M_0 \geq 0$) and becomes strictly positive beginning with a certain value $M_{0\mbox{\tiny min}}$. Thus, any solution of Eq.\,(\ref{gap2}) with $M_0>M_{0\mbox{\tiny min}}$ gives $\delta_M>0$. As we will show later, the value of $\delta_M$ is uniquely determined in the model, but first we should discuss the role of Daschen's theorem in the parameter-fixing procedure. 

Let us return to mass formulas. From (\ref{pi1})-(\ref{K01}) it follows that 
\begin{eqnarray}
\label{udFull}
&&\frac{\bar m_{K^+}^2\!-\bar m_{K^0}^2\!+\bar m_{\pi^+}^2}{\bar m_{K^0}^2\!-\bar m_{K^+}^2\!+\bar m_{\pi^+}^2} 
= \frac{m_u}{m_d}\! -\! \frac{m_s\delta_M}{2M_0}\left(\!1\!-\!\frac{m_u^2}{m_d^2}\right)\!, \\
\label{sdFull}  
&&\frac{\bar m_{K^+}^2\!+\bar m_{K^0}^2\!-\bar m_{\pi^+}^2}{\bar m_{K^0}^2\!-\bar m_{K^+}^2\!+\bar m_{\pi^+}^2} 
= \frac{m_s}{m_d}\! +\! \frac{m_u\delta_M}{2M_0}\left(\frac{m_s^2}{m_d^2}\!-\!1\!\right)\!.
\end{eqnarray}  
The current algebra result arises from here in the leading order of the chiral expansion,  
\begin{eqnarray}
\label{udchex}
&& \frac{m_u}{m_d}=\frac{\bar \mu_{K^+}^2 -\bar \mu_{K^0}^2+\bar \mu_{\pi^+}^2}{\bar \mu_{K^0}^2-\bar \mu_{K^+}^2+\bar \mu_{\pi^+}^2}\equiv R_x, \\
\label{sdchex}  
&&\frac{m_s}{m_d}=\frac{\bar \mu_{K^+}^2+\bar \mu_{K^0}^2-\bar \mu_{\pi^+}^2}{\bar \mu_{K^0}^2-\bar \mu_{K^+}^2+\bar \mu_{\pi^+}^2}\equiv R_y.
\end{eqnarray} 

Additionally, one may wish to take into account the electromagnetic interaction of charged particles, which increases the masses of these states:
\begin{eqnarray}
  \mu_{\pi^+}^2 &=&\bar\mu^2_{\pi^+} +\Delta^2_{el}, \quad \mu^2_{\pi^0}=\bar\mu^2_{\pi^0}=\bar\mu^2_{\pi^+}, \\
  \mu_{K^+}^2&=&\bar\mu^2_{K^+} +\tilde\Delta^2_{el}, \quad \mu^2_{K^0}=\bar\mu^2_{K^0}.
\end{eqnarray} 
The difference between the masses of the charged and neutral pions $\mu_{\pi^+}>\mu_{\pi^0}$ is due primarily to the electromagnetic interaction. The contribution of the strong interaction is proportional to $(m_d-m_u)^2$ and is thereby negligibly small. Using the Dashen theorem  \cite{Dashen:69}
\begin{equation}
\label{Dteor}
  \Delta^2_{el}=\tilde\Delta^2_{el},
\end{equation}
which is a strict result of the current algebra, one arrives at the well-known Weinberg ratios \cite{Weinberg:77}
\begin{eqnarray}
 \label{ud} 
   \frac{m_u}{m_d}&=&\frac{2\mu^2_{\pi^0}-\mu^2_{\pi^+}+\mu^2_{K^+}-\mu^2_{K^0}}{\mu^2_{K^0}-\mu^2_{K^+}+\mu^2_{\pi^+}}=0.56, \\
  \label{sd}
  \frac{m_s}{m_d}&=&\frac{\mu^2_{K^+}+\mu^2_{K^0}-\mu^2_{\pi^+}}{\mu^2_{K^0}-\mu^2_{K^+}+\mu^2_{\pi^+}}=20.18. 
\end{eqnarray}
Taking into account the first $1/N_c$ correction in (\ref{udFull})-(\ref{sdFull}) and after the inclusion of electromagnetic corrections, we arrive to the Leutwyler inequalities  \cite{Leutwyler:96a}
\begin{eqnarray}
 \label{Lud} 
  \frac{m_u}{m_d}&>&\frac{2m^2_{\pi^0}-m^2_{\pi^+}+m^2_{K^+}-m^2_{K^0}}{m^2_{K^0}-m^2_{K^+}+m^2_{\pi^+}}\equiv R_{xD}, \\
  \label{Lsd}
  \frac{m_s}{m_d}&<&\frac{m^2_{K^+}+m^2_{K^0}-m^2_{\pi^+}}{m^2_{K^0}-m^2_{K^+}+m^2_{\pi^+}}\equiv R_{yD}, 
\end{eqnarray}
which are valid at $\delta_M > 0$ (here and below, the subscript $D$ marks expressions that are derived with the Dashen theorem). If $\delta_M < 0$, the reverse inequalities are fulfilled. 

The specific case $\delta_M=0$ indicates that the first correction to the result of the current algebra vanishes and the Weinberg ratios are satisfied, which is possible if $M_0 = M_{0\mbox{\tiny min}}$. In this case, at $\Lambda = 1.1\,\mbox{GeV}$, the rest six parameters $G_S, G_V$, $m_u$, $m_d$, $m_s$ and $\Delta^2_{el}$ can be fixed by the phenomenological values of the masses $m_{\pi^0}$, $m_{\pi^+}$, $m_{K^0}$, $m_{K^+}$, the weak decay constant of the pion $f_\pi = 92.3\pm 0.1\,\mbox{MeV}$, and the requirement that the first order correction to the current algebra result is suppressed $\delta_M = 0$ (in this case the Dashen theorem is exact). As a result, we obtain $M_0 = M_{0\mbox{\tiny min}} = 236\,\mbox{MeV}$, $G_S = 6.35\,\mbox{GeV}^{-2}$, $G_V = 4.28\,\mbox{GeV}^{-2}$; the magnitude of the quark condensate is $\langle 0|\bar qq|0\rangle_0 = -(265\,\mbox{MeV})^3$; the light quark masses are $m_u = 2.8\,\mbox{MeV}$, $m_d=5.0\,\mbox{MeV}$, and $m_s =101\,\mbox{MeV}$. The parameter characterizing the relative magnitude of breaking of isotopic symmetry compared to the breaking of $SU(3)_f$ symmetry is
\begin{equation}
\label{R}
R=\frac{m_s-\hat m}{m_d-m_u}=44.
\end{equation}  
All these results are summarized in the Table\,\ref{ParameterSets} (see set (a) with $\delta_M=0$). 

The numerical values given in Table\,\ref{ParameterSets} for the set (a) appear to be reasonable in the $u$ and $d$ sector. However, in the strange sector the NLO correction $M_1(m_s)=432\,\mbox{MeV}$ is almost twice as large as
the LO result $M_0=236\,\mbox{MeV}$. This strongly violates the main condition of the asymptotic expansion (\ref{AS}) and means that our assumption $\delta_M=0$ fails for strange quark. Thus we are led to conclude that $\delta_M>0$. As we already mentioned above, Leutwyler \cite{Leutwyler:96a} obtained a similar inequality $\Delta_M>0$ (both parameters are related through the Eq.\,(\ref{DeM})), from the in-depth analysis of the spectrum of neutral states $\pi^0$, $\eta$, $\eta'$ with the inclusion of effects caused by the breaking of $U(1)_A$ symmetry and the Zweig rule and classified it as a generous lower bound for the region where a truncated $1/N_c$ expansion leads to meaningful results. We come to a similar conclusion based solely on the NLO solution of the gap equation.

\begin{table*}
\caption{The six parameters of the model $\Lambda$, $G_{S,V}$, $m_{u,d,s}$ and electromagnetic correction to the masses of charged mesons, $\Delta^2_{el}$, $\tilde\Delta^2_{el}$, are fixed by using the meson masses $m_{\pi^0}$, $m_{\pi^+}$, $m_{K^0}$, $m_{K^+}$, the weak pion decay constant $f_\pi$ and the cutoff $\Lambda$ as an input (input values are marked with an asterisk ($^\ast$)). In the first row of the table, set $(a)$, the condition $\delta_M=0$ is used as the seventh input value. This set describes a hypothetical case of the complete absence of the first $1/N_c$ correction to the mass formulas (\ref{pi1})-(\ref{K01}). Set $(b)$ describes a realistic case when the first correction is nonzero (here $f_K$ and $\eta\to 3\pi$ decay rate are used as additional input values). All units, except $\left[G_{S,V}\right]=\text{GeV}^{-2}$ and dimensionless ratio $R$ (see Eq.\,(\ref{R})), are given in MeV.}
\label{ParameterSets}
\begin{footnotesize}
\begin{tabular*}{\textwidth}{@{\extracolsep{\fill}}lrrrrrrrrcrrrrrcr@{}}
\hline
\hline 
\multicolumn{1}{c}{Set}
& \multicolumn{1}{c}{$\delta_M$}
& \multicolumn{1}{c}{$\Lambda$} 
& \multicolumn{1}{c}{$G_S$}
& \multicolumn{1}{c}{$G_V$}
& \multicolumn{1}{c}{$m_u$}
& \multicolumn{1}{c}{$m_d$}
& \multicolumn{1}{c}{$m_s$}
& \multicolumn{1}{c}{$M_0$}
& \multicolumn{1}{c}{$-\langle\bar qq\rangle^{1/3}_0$}
& \multicolumn{1}{c}{$M_u$}
& \multicolumn{1}{c}{$M_d$}
& \multicolumn{1}{c}{$M_s$}
& \multicolumn{1}{c}{$F$}
& \multicolumn{1}{c}{$f_\pi$}
& \multicolumn{1}{c}{$f_K$}  
& \multicolumn{1}{c}{$R$} \\
\hline
$(a)$
& $0^\ast$ 
& $1.1^\ast$ 
& $6.35$  
& $4.28$ 
& $2.8$ 
& $5.0$ 
& $101$ 
& $236$ 
& $265$ 
& $248$   
& $257$  
& $668$    
& $89.0$  
& $92.2^\ast$  
& $131$ 
& $44$\\
$(b)$
& $ 0.67$  
& $1.1^\ast$ 
& $6.6$  
&  $7.4$ 
& $2.6$ 
& $4.6$ 
& $84$ 
& $274$ 
& $275$ 
& $283$   
& $290$  
& $567$    
& $90.5$  
& $92.2^\ast$  
& $111^\ast$ 
& $40$\\
\hline
\hline
\end{tabular*}
\end{footnotesize} 
\end{table*}

If the first correction is nonzero and $M_0>M_{0\mbox{\tiny min}}$, i.e., $\delta_M>0$, the value of the quark condensate increases. Consequently, the light quark masses decrease. Therefore, the above estimates for the masses $m_u$, $m_d$, and $m_s$ should be considered as an upper bound of the model.

\section{The Gasser-Leutwyler ellipse and higher order curves}
\label{s7}
Let us return to the analysis of mass formulas (\ref{pi1})-(\ref{K01}) at $\delta_M \neq 0$. A number of sum rules which do not involve $\delta_M$ can be obtained from these formulas. 

Gasser and Leutwyler \cite{Gasser:85} considered the simplest case described by a second-order curve in two independent variables $x = m_u/m_d$ and $y = m_s/m_d$. We arrive at this curve using two ratios
\begin{eqnarray}
  \label{Ratio1}
  &&R_1\!=\!\frac{\bar  m_{K^+}^2}{\bar m_{\pi^+}^2}\!=\!\frac{m_u\!+\!m_s}{m_u\!+\!m_d}\left[1\!+\!\frac{m_s\!-\!m_d}{2M_0} \,\delta_M\right]\!,  \\
  \label{Ratio2}
  &&R_2\!=\!\frac{\bar m_{K^0}^2\!-\!\bar  m_{K^+}^2}{\bar m_{K^0}^2\!-\!\bar m_{\pi^+}^2}\!=\!\frac{m_d\!-\!m_u}{m_s\!-\!m_u}\left[1\!+\!\frac{m_s\!-\!m_d}{2M_0} \,\delta_M\right]\!,
\end{eqnarray}     
from which it follows that
\begin{equation}
\label{Q2}
Q^2\equiv \frac{R_1}{R_2} =\frac{m_s^2-m^2_u}{m_d^2-m_u^2}.
\end{equation}  
The right hand side of this expression depends only on the ratios $x$ and $y$ of the light quark masses. The locus of these points is an ellipse
\begin{equation}
\label{ellipse}
y^2-x^2(1-Q^2)=Q^2. 
\end{equation}  

Note that replacing $\bar m_{K^+}\!\leftrightarrow\! \bar m_{K^0}$ in the left-hand side of equations Eqs.\,(\ref{Ratio1}) and (\ref{Ratio2}) reduces to replacing $m_u\!\leftrightarrow\! m_d$ in the right-hand side, and it would seem that we arrive at a new relation
\begin{equation}
\label{Q22}
\tilde Q^2\equiv \frac{\bar  m_{K^0}^2}{\bar  m_{\pi^+}^2}\frac{\bar  m_{K^+}^2-\bar  m_{\pi^+}^2}{\bar  m_{K^0}^2-\bar  m_{K^+}^2} =\frac{m_s^2-m^2_d}{m_d^2-m_u^2}.
\end{equation}  
However, it is easy to see that this equation coincides with (\ref{Q2}) because $Q^2=\tilde Q^2+1$.

Taking into account electromagnetic corrections, according to the Dashen theorem, we obtain an ellipse with a semimajor axis $Q \to Q_D$
\begin{eqnarray}
\label{QR12D}
Q^2\to Q^2_D&=&\frac{(m_{K^0}^2-m_{\pi^0}^2)(m_{K^+}^2-m_{\pi^+}^2+m_{\pi^0}^2)}{m_{\pi^0}^2 (m_{K^0}^2-m_{K^+}^2+m_{\pi^+}^2-m_{\pi^0}^2)}, \nonumber \\
R_1\to R_{1D}&=&1+\frac{m_{K^+}^2-m_{\pi^+}^2}{m_{\pi^0}^2},  \\
R_2\to R_{2D}&=&1-\frac{m_{K^+}^2-m_{\pi^+}^2}{m_{K^0}^2-m_{\pi^0}^2}, \nonumber
\end{eqnarray}   
which gives $Q_D = 24.3$, $R_{1D}=13.3$, $R_{2D}=0.0225$ for physical values of the masses. Obviously, the point $(x, y) = (R_{xD}, R_{yD})$ belongs to this ellipse. Below, for the sake of brevity, this point is called the Weinberg point, where $\delta_M = 0$ and, consequently, the Weinberg ratios (\ref{ud}) and (\ref{sd}) are satisfied. 

Consider now a set of arbitrary ratios $R_{i}$ ($i =1,2,\ldots$) combined from the meson masses (\ref{pi1})-(\ref{K01})
\begin{equation}
\label{Ri}
R_i=k_i\left(1+l_i\frac{m_d}{2M_0}\delta_M\right), 
\end{equation}
where coefficients $k_i$ and $l_i$ are functions of $x$ and $y$. It is clear that by taking two arbitrary elements of the given set, say $R_i$ and $R_j$, we can eliminate the dependence on $m_d\delta_M/2M_0$ and arrive at the equation of the curve in the $(x,y)$ plane
\begin{equation}
\label{gencase1}
k_ik_j(l_i-l_j)=l_ik_iR_j-l_jk_jR_i.
\end{equation}
If $l_i=l_j$, the equation simplifies to $k_iR_j=k_jR_i$. This curve, where $i=1,j=2$, is an ellipse (\ref{ellipse}). If $l_i\neq l_j$, the curve (\ref{gencase1}) is of a higher order, and consequently the allowed values of $x$ and $y$ do not belong to the ellipse. Hence we have a set of alternative sum rules to determine the light quark mass ratios. The ellipse is distinguished by two properties: (i) It survives even after accounting for chiral logarithms, (ii) It has an additional symmetry with respect to replacement $m_u\!\leftrightarrow\! m_d$. Nevertheless, in the approximation considered here, it is difficult to give preference to one of curves. Let us describe the most important property of the family.  

\begin{figure}
\includegraphics[width=0.45\textwidth]{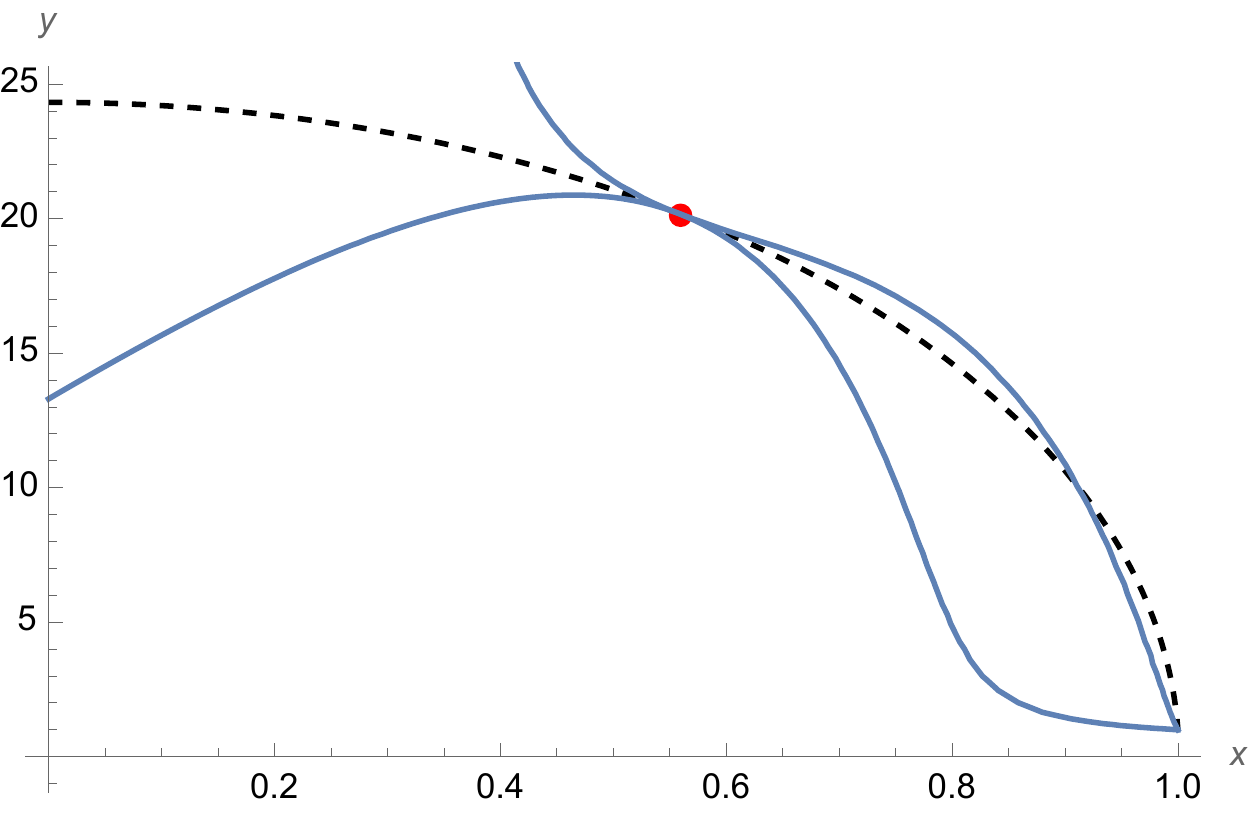}
\caption{Ellipse (dashed line) specified by Eq.\,(\ref{ellipse}), where $Q\to Q_D$, and curves given by  Eq.\,(\ref{lbc}) (lower solid line) and Eq.\,(\ref{ubc}) (upper solid line) obtained with the Dashen theorem ($R_{1,3,x}\to R_{1D,3D,xD}$). The dot belonging to all three curves is the Weinberg point. The curves intersect at the $SU(3)$ limit point $(x,y)=(1,1)$.}
\label{fig2}      
\end{figure}

For this purpose, note that the family (\ref{gencase1}) is enclosed between two curves. The first one is given by the ratios $R_1$ (see Eq.\,(\ref{Ratio1})) and $R_3$
\begin{eqnarray}
 \label{us} 
   R_3&=&\frac{\bar m^2_{K^+}+\bar m^2_{K^0}-\bar m^2_{\pi^+}}{\bar m^2_{K^+}-\bar m^2_{K^0}+\bar m^2_{\pi^+}}\nonumber \\
   &=&\frac{y}{x}\left[1+\left(\frac{y}{x}-\frac{x}{y}\right)\frac{m_d\delta_M}{2M_0}\right]. 
\end{eqnarray}
In this specific case, Eq.\,(\ref{gencase1}) leads to the elliptic curve 
\begin{equation}
\label{lbc}
x(y-1)(y-xR_3)=(x-y)(1+x)R_1+y^2-x^2
\end{equation}
which has two connected components one of which passes through the Weinberg point and determines the lower bound in Fig.\,\ref{fig2}. To plot the curve (\ref{lbc}) we use the physical values of the meson masses, where electromagnetic corrections are taken into account in accord with the Dashen theorem; i.e., $R_1\to R_{1D}$, and
\begin{equation}
R_3\to R_{3D}=\frac{m^2_{K^+}+m^2_{K^0}-m^2_{\pi^+}}{2m^2_{\pi^0}+m^2_{K^+}-m^2_{K^0}-m^2_{\pi^+}}.
\end{equation}

To establish the upper bound of the family (\ref{gencase1}) we consider the ratios $R_3$ and $R_x$. It leads to the fifth order curve
\begin{equation}
\label{ubc}
xy(1-x^2)(xR_3-y)=(y^2-x^2)(x-R_x).
\end{equation}
It has three connected components. Fig.\,\ref{fig2} shows the component passing through the Weinberg point. It is also obtained with the Dashen theorem: $R_3\to R_{3D}$ and $R_x \to R_{xD}$, and it lies primarily above the ellipse specified by Eq.\,(\ref{ellipse}). The other curves lie inside indicated boundaries. The common property of the family is that all of them pass through the Weinberg point. The existence of numerous curves generated by mass formulas (\ref{pi1})-(\ref{K01}) does not affect the Leutwyler inequalities (\ref{Lud}) and (\ref{Lsd}). The question which of the curves (sum rules) is more suitable for approximating the final result can be clarified only after the model parameters are fixed.

\section{Numerical estimates}
\label{s8}
Let us fix the six parameters of the model $\Lambda$, $G_S$, $G_V$, $m_u$, $m_d$, $m_s$,   and electromagnetic corrections to the masses of charged mesons $\Delta_{el}$ and $\tilde\Delta_{el}$. For a direct comparison with the empirical data and $1/N_c\chi$PT results, we have used the values for the pion and kaon decay constants, $f_\pi\simeq 92\,\mbox{MeV}$, $f_K\simeq 110\,\mbox{MeV}$, and the masses of pseudoscalar mesons: $m_{\pi^0}$, $m_{\pi^+}$, $m_{K^0}$, $m_{K^+}$. Let us recall that $\Delta_{el}^2\simeq m^2_{\pi^+}-m^2_{\pi^0}$. In addition, as noted above, we choose the cutoff $\Lambda$ according to a generally accepted estimate of the scale of spontaneous chiral symmetry breaking $\Lambda_{\chi SB}\simeq 4\pi f_\pi$ \cite{Georgi:84}. To fix $\tilde\Delta_{el}$ we use (following the Leutwyler's analyses \cite{Leutwyler:97}) the $\eta\to 3 \pi$ decay. The latter requires some clarifications. 

\begin{figure}
\includegraphics[width=0.45\textwidth]{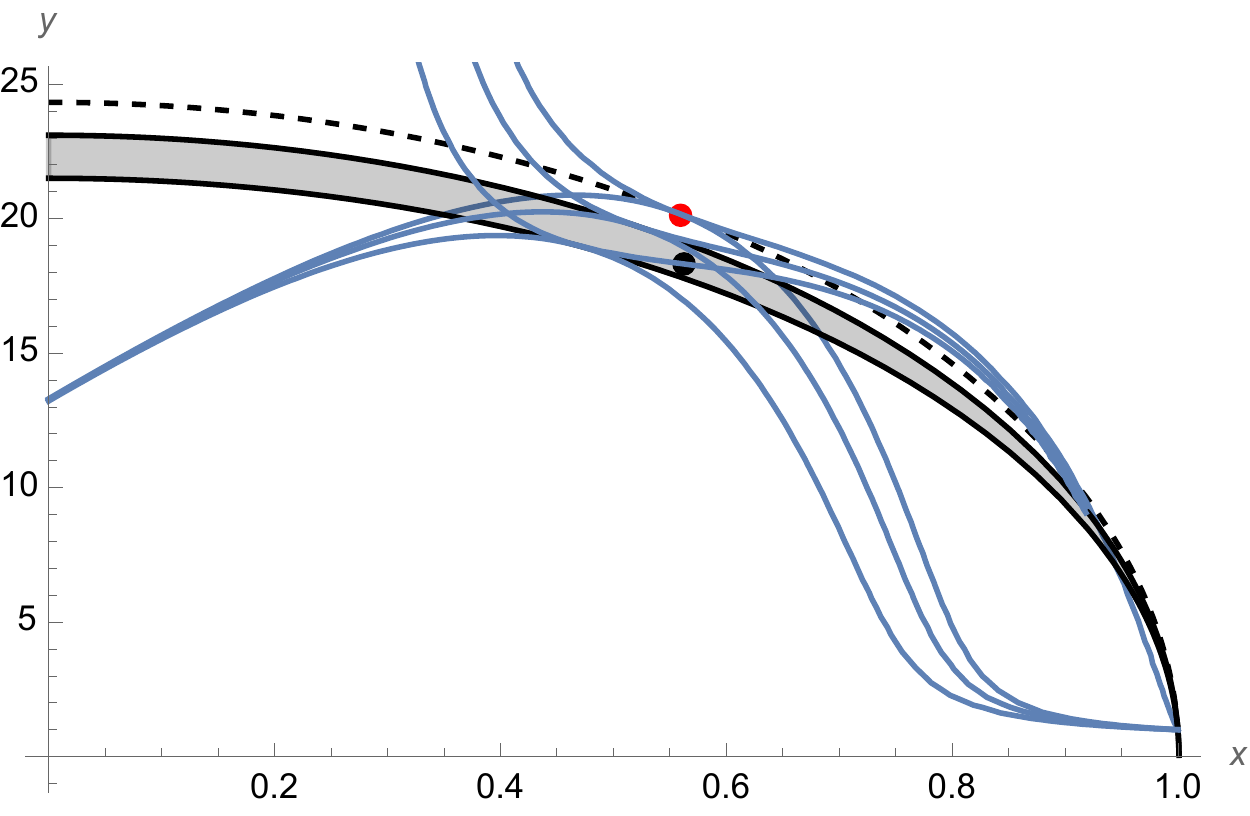}
\caption{The area of admissible values of the ratios $x, y$ is formed by the intersection of three bands corresponding to the interval $\tilde\Delta_{el}=(47.1\mp 4.5)\,\mbox{MeV}$ at the ellipse (gray band) and at curves of higher order (\ref{lbc}) and (\ref{ubc}). The dot belonging to all three bands is the estimate of the NJL model.}
\label{fig3}      
\end{figure}

The Dashen theorem is valid only in the leading order of chiral expansion. Since $\delta_M\neq 0$, the Eq.\,(\ref{Dteor}) no longer holds. The inequality $\tilde\Delta_{el}^2\neq \Delta_{el}^2=m_{\pi^+}^2-m_{\pi^0}^2$ should be considered instead. It follows that the ratio (\ref{Q2}) depends now on $\tilde \Delta_{el}^2$
\begin{equation}
Q^2_{\tilde D}=\left(\frac{m_{K^0}^2}{m_{\pi^0}^2}-1\right)\left(\frac{m_{K^0}^2}{m_{K^0}^2-m_{K^+}^2+\tilde\Delta_{el}^2}-1\right)\!.
\end{equation}
In \cite{Leutwyler:97} it was argued that the value of $Q^2_{\tilde D}$ can be extracted accurately from the observed $\eta\to 3 \pi $ decay width. The reason is that electromagnetic contributions to this process are suppressed, and, as a consequence, determination of $Q^2_{\tilde D}$ is less sensitive to the uncertainties therein. The current knowledge based on $\eta\to 3 \pi $ gives the range $Q_{\tilde D} = 22.3 \pm 0.8 $ \cite{Leutwyler:09}, which leads to $\tilde\Delta_{el}=(47.1\mp 4.5)\,\mbox{MeV}$ correspondingly. For comparison, $\Delta_{el} = 35.5\,\mbox{MeV}$. The gray elliptic band in Fig.\,\ref{fig3} corresponds to the range $\tilde\Delta_{el}$ indicated above. Similar bands are also plotted for the higher order curves. Obviously, the closer the value of $\tilde\Delta_{el}$ to $\Delta_{el}$, the closer the curve is to the corresponding one obtained taking into account Dashen's theorem.  

The interval we use slightly differs from the recent result $Q_{\tilde D} =22.1\pm 0.7$ \cite{Colangelo:18}. Nonetheless, we prefer to consider more wider region because the lattice QCD collaborations report on larger values: $Q_{\tilde D}=23.4\pm 0.6$ \cite{Fodor:16} for $N_f=2+1$ and $Q_{\tilde D}=23.8\pm 1.1$ \cite{Giusti:17} for $N_f=2+1+1$ simulations. 

The above estimate for $\tilde\Delta_{el}$ means that the mass difference between charged and neutral kaons due to electromagnetic interactions is $(m_{K^+}-m_{K^0})_{el}=(2.2\mp 0.4)\,\mbox{MeV}$. It agrees with the result of lattice QCD calculations $(m_{K^+}-m_{K^0})_{el}=1.9\,\mbox{MeV}$ \cite{Duncan:96}. 

The input values give the following estimates for the couplings $G_S=6.6\,\mbox{GeV}^{-2}$ and $G_V=7.4\,\mbox{GeV}^{-2}$. These constants, in particular, describe the theory in the limit $N_c\to\infty$, i.e., when the masses of the current quarks vanish and $f_\pi=f_K=F$. This means that their values should mainly determine some vacuum characteristics. Indeed, after fixing the parameters, we see that
\begin{eqnarray}
\sqrt{\frac{1}{2G_S}}&\simeq& M_0\simeq |\langle \bar qq\rangle_0^{1/3}|, \nonumber \\
\sqrt{\frac{Z_0-1}{4G_VZ_0}}&\simeq&\sqrt{\frac{1}{16G_V}}\simeq F.  
\end{eqnarray}
Some of numerical estimates are given in the second line of the Table\,\ref{ParameterSets}. For $Z_0$ we have $Z_0=1.32$, taking into account the $1/N_c$ correction this gives $Z_\pi =1.34$, $Z_K=1.51$, which if averaged coincides with the estimate of the standard NJL model $Z=1.4$ \cite{Volkov:86}. 

The Gell-Mann, Oakes and Renner result for $B_0$ is modified at NLO by the factor, which is the ratio of the squares of the masses of the pseudoscalar meson, to its value at LO, i.e., 
\begin{equation}
B_0\to B_P=B_0 \left(\frac{\bar m_P^2}{\bar\mu_P^2}\right),
\end{equation}
where $P=\pi^\pm, K^\pm, K^0$-$\bar K^0$. Numerically the correction is less than 1\% for pions, and around 11\% for kaons. Thus, the correction of $\mathcal O(m^2_i)$ in the $1/N_c$ expansion of the pseudoscalar masses is much less than the LO result. This supports the assumption made that the quark condensate is of the order $N_c$.

Mass formulas (\ref{pi1})-(\ref{K01}) allow to obtain the absolute values of the quark masses, if one knows the vacuum characteristics encoded in the parameter $B_0$, and in the ratio $\delta_M/M_0$. We need also to take into account the electromagnetic corrections. Given that the difference of charge and neutral pion masses has mainly electromagnetic origin, one finds 
\begin{eqnarray}
&&\bar m_{\pi^\pm}^2=m_{\pi^\pm}^2-\Delta^2_{el}=m_{\pi^0}^2, \nonumber \\
&&\bar m_{K^\pm}^2=m_{K^\pm}^2-\bar\Delta_{el}^2, \nonumber \\
&&\bar m_{K^0}^2=m_{K^0}^2,
\end{eqnarray}
where $m_{\pi^0}$, $m_{\pi^\pm}$, $m_{K^\pm}$, and $m_{K^0}$ are the physical masses of the states. We collect our results in the first line of the Table \ref{QMR}. The error bars there indicate the change in values within the range $\tilde\Delta_{el}=(47.1\mp 4.5)\,\mbox{MeV}$. The results of calculations in $1/N_c\chi$PT \cite{Leutwyler:96} (second line) and data quoted by the Particle Data Group (PDG) \cite{PDG:22} (third line) are also given there.

\begin{table*}
\caption{The light quark masses (in MeV) and their ratios obtained in the $1/N_c$ NJL model are compared with the results of $1/N_c\chi$PT and PDG. In the first line, the error bars indicate the change in values within the range $\tilde\Delta_{el}=(47.1\mp 4.5)\,\mbox{MeV}$ correspondingly. }
\label{QMR}
\begin{footnotesize}
\begin{tabular*}{\textwidth}{@{\extracolsep{\fill}}lllllllll@{}}
\hline
\hline 
\multicolumn{1}{c}{Set}
& \multicolumn{1}{c}{$m_u $}
& \multicolumn{1}{c}{$m_d$}
& \multicolumn{1}{c}{$m_s$}
& \multicolumn{1}{c}{$m_u/m_d$}
& \multicolumn{1}{c}{$m_s/m_d$} 
& \multicolumn{1}{c}{$m_s/\hat m$}
& \multicolumn{1}{c}{$m_s/m_u$} 
& \multicolumn{1}{c}{$R$} \\
\hline
$1/N_c \mbox{NJL}$
& $2.57\pm 0.07$
& $4.56^{-0.06}_{+0.08}$
& $83.7\pm 0.1$
& $0.564^{-0.025}_{+0.023}$ 
& $18.3 \mp 0.3$ 
& $23.46\mp 0.02$  
& $32.5\mp 0.9$ 
& $40.3^{+2.9}_{-2.8}$ 
\\
$1/N_c \chi\mbox{PT}$ \cite{Leutwyler:96}
& 
& 
& 
& $0.553\pm 0.043$  
& $18.9\pm 0.8$ 
& $24.4\pm 1.5$  
&  $34.4\pm 3.7$ 
&  $40.8\pm 3.2$
\\
$\mbox{PDG}$ \cite{PDG:22}
& $2.16^{+0.49}_{-0.26}$
& $4.67^{+0.48}_{-0.17}$
& $93.4^{+8.6}_{-3.4}$
& $0.474^{+0.056}_{-0.074}$  
& $19.5\pm 2.5$ 
& $27.33^{+0.67}_{-0.77}$  
&  $$ 
&  $$
\\
\hline
\hline
\end{tabular*}
\end{footnotesize} 
\end{table*}

We can conclude that the results of the $1/N_c$ NJL model are in a remarkable agreement with the estimates made in \cite{Leutwyler:96}. This is also evidenced by the estimates we obtain for a number of parameters of the $1/N_c$ chiral perturbation theory. For the problem considered here these are the following constants: $L_5^r$, $L_8^r$, $\Delta_M$, $R$. Our estimates are
\begin{equation}
L_5^r= 2.1\cdot 10^{-3}, \quad
L_8^r= 1.3\cdot 10^{-3}, \quad
\Delta_M= 0.10.
\end{equation}    
This agrees with the the phenomenological estimates of these low-energy coupling constants $L_5^r= (2.2\pm 0.5) \cdot 10^{-3}$ and $L_8^r= (1.1\pm 0.3)\cdot 10^{-3}$ in \cite{Gasser:85}. The parameter $\Delta_M$ cannot be calculated within $1/N_c\chi$PT. Our result is compatible with the estimate $0 < \Delta_M \leq 0.13$ \cite{Leutwyler:96a}. For the ratio $R$ the $1/N_c\chi$PT yields $R=40.8\pm 3.2$. Our calculations give $R=40.3^{+2.9}_{-2.8}$. Let us also indicate the independent estimate $R = 41\pm 4$ obtained in the work \cite{Urech:95}.

Evaluation of the absolute values of the current quark masses, shows that our result differs little from the standard estimates obtained in the NJL model. For comparison, let us point out the paper \cite{Volkov:86}, where it was found that $m_u=m_d=3\,\mbox{MeV}$ and $m_s=80\,\mbox{MeV}$. Recall that the resulting masses only make sense in the limited context of a particular quark model, and cannot be related to the quark mass parameters of the Standard Model. However, the quark-mass ratios found here have a broader meaning because they do not depend on the absolute normalization of the quark mass.

The point in the Fig.\,\ref{fig3} inside the gray band corresponds the central value of the ratios $m_u/m_d$ and $m_s/m_d$ presented by the set $1/N_c$\,NJL in the Table \ref{QMR}. 

It is also interesting to note that the bounds on the two ratios $m_u/m_d$ and $m_s/m_d$ can be established solely on the sum rules (\ref{gencase1}). Since all curves (\ref{gencase1}) intersect at a Weinberg point, considering only the extreme curves automatically cuts out the area belonging to the entire curve family in question. In Fig.\,\ref{fig3} we show two bands obtained on the bases of the extremal curves (\ref{lbc}) and (\ref{ubc}), from their intersection one can deduce that 
\begin{equation}
\label{curmassratios-extr}
\frac{m_u}{m_d}=0.50\pm 0.09, \quad \frac{m_s}{m_d}=19.22 \pm  0.62. 
\end{equation}
These ratios do not imply any definite value for the parameter $\delta_M$ and therefore can be considered as the maximum range for possible values $m_u/m_d$ and $m_s/m_d$ that arises in the truncated theory based on formulas (\ref{pi1})-(\ref{K01}). This result agrees well with the values $m_u/m_d=0.474^{+0.056}_{-0.074}$ and $m_s/m_d=19.5\pm 2.5$ quoted by the Particle Data Group \cite{PDG:22}.  

A narrower interval is obtained if we additionally require the fulfillment of the Leutwyler's inequalities (\ref{Lud}) and (\ref{Lsd}), i.e., $\delta_M>0$. In this case, the result reads  
\begin{equation}
\label{curmassratios-extr2}
\frac{m_u}{m_d}=0.53\pm 0.06, \quad \frac{m_s}{m_d}=19.13 \pm  0.53. 
\end{equation}

Since we are dealing with a non-renormalizable model, the results depend on the regularization scheme used. Regularization is one of the elements that make up the basis of the NJL model, and different versions of the model differ in its choice. This aspect of NJL model is reviewed in detail by Klevansky \cite{Klevansky:92}. The proper-time regularization, which we use here, has also been carefully analyzed in \cite{Cvetic:97}. Nevertheless, we would like to dwell on one property of the gap equation, which should be kept in mind when comparing results of calculations with experimental data. The solution of the gap equation (\ref{gap2}), $M_0(\Lambda)$, changes drastically with $\Lambda$ at fixed value of $G_S$. Indeed, one easily finds that $M_0 (1.0\,\mbox{GeV}) =19.7\, \mbox{MeV}$, $M_0 (1.1\,\mbox{GeV}) =274\, \mbox{MeV}$, $M_0 (1.2\,\mbox{GeV}) =468\, \mbox{MeV}$ for $G_S=6.6\,\mbox{GeV}^{-2}$. Such behavior is associated with the original quadratic divergence of the integral $J_0(M_0)$, and is typical for any of the regularization schemes commonly used in the NJL model \cite{Klevansky:92}. As a consequence, the cut-off $\Lambda$ is sharply fixed, because even a $1\%$ change in the value of the cutoff $\Lambda$ leads to about $7\%$ change in the value of $M_0$, namely
\begin{equation}
\Lambda = 1.10\pm 0.01\,\mbox{GeV}, \quad M_0=274\pm 20\,\mbox{MeV}.
\end{equation}
This observation is crucial for a systematic study of the theoretical uncertainties associated with the regularization scheme used in the NJL model. The reason is that all other vacuum characteristics somehow depend on these two parameters and $G_V$. In particular, we find $a=3.50^{-0.28}_{+0.33}$, $\delta_M=0.67^{+0.10}_{-0.12}$, $F=90.54^{+2.82}_{-3.10}\,\mbox{MeV}$ and $|\langle \bar qq\rangle_0^{1/3} |=275^{+6}_{-7}\,\mbox{MeV}$. The above estimates make it possible to understand the magnitude of the theoretical uncertainties associated with the regularization scheme we apply.

The property of the solution of the gap equation mentioned above has a significant effect not only on the value of $M_0$, but also on the value of the first correction to the masses of the constituent quarks, which is proportional to the derivative $M'(0)=a$ entering the truncated formula (\ref{expm}). The mass of the strange quark $M_s=567\,\mbox{MeV}$ indicated in Table \ref{ParameterSets} (see the set (b)), follows from (\ref{expm}). Although this result agrees with the estimate $M_s=522.2\,\mbox{MeV}$ \cite{Klimt:90}, obtained by numerically solving the gap equation in the standard approach, there may be doubts about the selfconsistensy of our calculations, since the condition $M_0 \gg M_1$ is not satisfied. Formally, it certainly holds for $N_c\to\infty$, but for $N_c=3$ the NLO correction $M_1=a m_s=293\,\mbox{MeV}$ to the LO result, $M_0=274\,\mbox{MeV}$, turns out to be comparable with it. The pictures emerging at $N_c=\infty$ and $N_c=3$ seem to be quantitatively different from each other. Does this mean that the $1/N_c$ expansion is inapplicable in the case of the s-quark? Of course not. There are a number of similar examples \cite{Schifman:81}. Such a numerical deviation occurs when a large dimensional parameter penetrates into coefficients of the $1/N_c$ expansion. In the case considered, that is $B_0=2.5\,\mbox{GeV}$ -- a parameter associated with the quark condensate. It enters $a$ as
\begin{equation}
a=\frac{B_0}{2M_0}-\frac{G_V}{G_S}.
\end{equation}
Due to $B_0$ the pions and kaons are surprisingly heavy, given the light quark masses \cite{Shuryak:93}. The similar effect we observe in the mass of the strange constituent quark. Indeed, if one neglects the small mass of the up quark, the corresponding part of Eq.\,(\ref{expm}) can be written as 
\begin{equation}
M_s\simeq M_0+\frac{\bar\mu^2_{K^+}}{2M_0}\left(1-\frac{2G_VM_0}{G_SB_0}\right),
\end{equation}
i.e., the NLO correction to the strange quark mass is proportional to the kaon mass. PA-transitions reduce factor $1$ to $0.76$, which together with another factor $\bar\mu_{K^+}/(2M_0)$ determines the magnitude of the correction $M_1$. I should emphasize that the same mechanism is behind the violation of the $M_0\gg M_1$ condition at $N_c=3$ as behind the large value of the kaon mass. The latter problem should be explained by QCD.

\section{Conclusions} 
\label{Conclusions}
Here an attempt is made to implement, within the framework of the NJL model, the well established idea of Leutwyler, according to which the masses of light quarks are $1/N_c$ suppressed. This hypothesis previously turned out to be fruitful in constructing the effective Lagrangian of the $1/N_c\chi$PT. Extending this idea to the NJL model, we conclude that a coherent picture of masses and decay constants of electrically charged and strange pseudoscalars arises as well.    

Let us emphasize that in the NJL model, the vertices of the effective meson Lagrangian result from a direct calculation of the one-loop quark diagrams. This procedure needs to account for the effects of explicit chiral symmetry breaking. Many authors have taken steps in this direction. Still, we find that approach presented above leads to results and to a formulation of the situation that have not been yet reported in the literature (excluding a short letter published recently \cite{Osipov:22}). In particular, employing a recently developed method based on the Fock-Schwinger proper-time asymptotic expansion and the Volterra series, we demonstrate that this tool fairly reproduces the symmetry breaking pattern grasped by the effective Lagrangian of the $1/N_c\chi$PT.

The mass formulas obtained take into account the first $1/N_c$ correction to the result of current algebra. At this level it is possible to establish a set of mutually exclusive relations that directly relate the masses of $\pi^{\pm}$, $K^\pm$, $K^0$ and $\bar K^0$ mesons to the ratios of quark masses. We show that the $\eta\to 3\pi$ decay data do not allow giving preference to any one of the relations, but single out a physically significant range associated with all of them. The existence of such a range makes it possible to set limits for the light quark mass ratios $0.47<m_u/m_d<0.59$ and $18.60<m_s/m_d<19.66$, if one additionally requires that the Leutwyler inequalities are satisfied.

We have not considered here neutral $\pi^0$, $\eta$ and $\eta'$ states. They are the subject of a separate work, which is being prepared.

\section*{Acknowledgments}
I would like to thank B. Hiller for helpful correspondence and O.\,V. Teryaev and M.\,K. Volkov for useful conversations. This work is supported by Grant from Funda\c{c}\~ ao para a Ci\^ encia e Tecnologia (FCT) through the Grant No. CERN/FIS-COM/0035/2019. 



\begin{thebibliography}{99}
\bibitem{Hooft:74} G. 't Hooft, \emph{A planar diagram theory for strong interactions}, Nucl. Phys. B {\bf 72} (1974) 461-473.
\bibitem{Witten:79} E. Witten, \emph{Baryons in the $1/N$ expansion}, Nucl. Phys. B {\bf 160} (1979) 57-115.
\bibitem{Witten:80} E. Witten, \emph{Large N Chiral Dynamics}, Ann. of Phys. {\bf 128} (1980) 363-375.
\bibitem{Veneziano:80} P. Di Vecchia and G. Veneziano, \emph{Chiral dynamics in the large N limit}, Nucl. Phys. B {\bf 171} (1980) 253-272.
\bibitem{Trahern:80} C. Rosenzweig, J. Schechter and G. Trahern, \emph{Is the effective Lagrangian for quantum chromodinamics a $\sigma$ model?}, Phys. Rev. D {\bf 21} (1980) 3388-3392.
\bibitem{Ohta:80} K. Kawarabayashi and N. Ohta, \emph{The $\eta$ problem in the large-N limit: Effective Lagrangian approach}, Nucl. Phys. B {\bf 175} (1980) 477-492.
\bibitem{Ohta:81} K. Kawarabayashi and N. Ohta, \emph{On the partial conservation of the U(1) current},  Prog.  Theor. Phys. {\bf 66} (1981) 1709-1802.
\bibitem{Leutwyler:96a} H. Leutwyler, \emph{Bounds on the light quark masses}, Phys. Lett. B {\bf 374} (1996) 163-168.
\bibitem{Leutwyler:96b} H. Leutwyler, \emph{Implications of $\eta-\eta'$ mixing for the decay $\eta\to 3\pi$}, Phys. Lett. B {\bf 374} (1996) 181-185.
\bibitem{Taron:97} R. Herrera-Sikl\' ody, J.\,I. Latorre, P. Pascual, J. Taron, \emph{Chiral effective lagrangian in the large-$N_c$ limit: the nonet case}, Nucl. Phys. B {\bf 497} (1997) 345-386.
\bibitem{Kaiser:00} R. Kaiser and H. Leutwyler, \emph{Large $N_c$ in chiral perturbation theory}, Eur. Phys. J. C {\bf 17} (2000) 623-649. 
\bibitem{Osipov:06} A.\,A. Osipov, B. Hiller, J. da Provid\^ encia, \emph{Multi-quark interactions with a globally stable vacuum}, \emph{Phys. Lett. B} {\bf 634} (2006) 48-54. 
\bibitem{Weinberg:10} S. Weinberg, \emph{Pions in large N quantum chromodynamics}, Phys. Rev. Lett. {\bf 105} (2010) 261601. 
\bibitem{Witten:79b} E. Witten, \emph{Current algebra theorems for the $U(1)$ Goldstone boson}, Nucl. Phys. B {\bf 156} (1979) 269-283.
\bibitem{Veneziano:79} G. Veneziano, \emph{$U(1)$ without instantons}, Nucl. Phys. B {\bf 159} (1979) 213-224.
\bibitem{Goity:02} J. L. Goity, A. M. Bernstein and B. R. Holstein, \emph{Decay $\pi^0\to\gamma\gamma$ to next to leading order in chiral perturbation theory}, \emph{Phys. Rev. D} {\bf 66} (2002) 076014.
\bibitem{Bickert:20} P. Bickert and S. Scherer, \emph{Two-photon decays and transition form factors of $\pi^0$, $\eta$, and $\eta'$ in large-$N_c$ chiral perturbation theory}, \emph{Phys. Rev. D} {\bf 102} (2020) 074019.
\bibitem{Nambu:61a} Y. Nambu, G. Jona-Lasinio, \emph{Dynamical model of elementary particles based on an analogy with superconductivity. I}, \emph{Phys. Rev.} {\bf 122} (1961) 345-358.
\bibitem{Nambu:61b} Y. Nambu, G. Jona-Lasinio, \emph{Dynamical model of elementary particles based on an analogy with superconductivity. II}, \emph{Phys. Rev.} {\bf 124} (1961) 246-254.
\bibitem{Osipov:21a} A.\,A. Osipov, \emph{Fock-Schwinger method in the case of different masses}, \emph{JETP Letters} {\bf 113} No.6 (2021) 413-417. 
\bibitem{Osipov:21b} A.\,A. Osipov, \emph{Proper-time method for unequal masses}, \emph{Phys. Lett. B} {\bf 817} (2021) 136300.
\bibitem{Osipov:21c} A.\,A. Osipov, \emph{Proper-time evaluation of the effective action: Unequal masses in the loop}, \emph{Phys. Rev. D} {\bf 104} No.10 (2021) 105019.
\bibitem{Volkov:84} M.\,K. Volkov, \emph{Meson Lagrangians in a superconductor quark model}, \emph{Ann. of Phys.} {\bf 157}  (1984) 282-303.
\bibitem{Wadia:85} A. Dhar, R. Shankar, S.\,R. Wadia, \emph{Nambu--Jona-Lasinio--type effective Lagrangian: Anomalies and nonlinear Lagrangian of low-energy, large-$N$ QCD}, \emph{Phys. Rev. D} {\bf 31} (1985) 3256-3267.
\bibitem{Volkov:86} M.\,K. Volkov, \emph{Low energy physics of mesons in the superconducting quark model}, \emph{PEPAN} {\bf 17} (1986) 433-471.
\bibitem{Ebert:86} D. Ebert, H. Reinhardt, \emph{Effective chiral hadron Lagrangian with anomalies and Skyrme terms from quark flavour dynamics}, \emph{Nucl. Phys. B} {\bf 271} (1986) 188-226.
\bibitem{Bijnens:93} J. Bijnens, C. Bruno and E. de Rafael, \emph{Nambu--Jona-Lasinio-like models and the low-energy effective action of QCD}, \emph{Nucl. Phys. B} {\bf 390} (1993) 501-541.
\bibitem{Osipov:17} A.\,A. Osipov, M.\,K. Volkov, \emph{Chiral transformations of spin-1 mesons in the non-symmetric vacuum}, \emph{Ann. of Phys.} {\bf 382} (2017) 50-63.
\bibitem{Wess:69a} S. Coleman, J. Wess and B. Zumino, \emph{Structure of Phenomenological Lagrangians. I}, \emph{Phys. Rev.} {\bf 177} (1969) 2239-2247.
\bibitem{Wess:69b} C.\,G. Callan, S. Coleman, J. Wess and B. Zumino, \emph{Structure of Phenomenological Lagrangians. II}, \emph{Phys. Rev.} {\bf 177} (1969) 2247-2250.
\bibitem{Osipov:85} M.\,K. Volkov, A.\,A. Osipov, \emph{Decays of $B$, $H$, $H'$, $Q_1$ and $Q_2$ mesons in the quark model of the superconducting type}, \emph{Sov. J. Nucl. Phys.} {\bf 41} (1985) 500-503.
\bibitem{Klevansky:92} S.\,P Klevansky, \emph{The Nambu -- Jona-Lasinio model of quantum chromodynamics} \emph{Rev. Mod. Phys.} {\bf 64} (1992) 649-708.
\bibitem{Georgi:84} A. Manohar, H. Georgi, \emph{Chiral quarks and the non-relativistic quark model}, \emph{Nucl. Phys. B} {\bf 234} (1984) 189-212.
\bibitem{Ball:89} R.\,D. Ball, \emph{Chiral gauge theory}, \emph{Phys. Rep.} {\bf 182} (1989) 1-186.
\bibitem{Schwinger:51} J. Schwinger,  \emph{On gauge invariance and vacuum polarization}, \emph{Phys. Rev.} {\bf 82} (1951) 664-679.
\bibitem{DeWitt:65} B.\,S. DeWitt, \emph{Dynamical Theory of Groups and Fields}, Gordon \& Breach, New York (1965).
\bibitem{Feynman:51} R. P. Feynman, \emph{An operator calculus having applications in quantum electrodynamics}, \emph{Phys. Rev.} {\bf 84} (1951) 108-128.
\bibitem{Styan:73} G. P. H. Styan, \emph{Hadamard products and multivariate statistical Analysis}, \emph{Linear. Algebra and its Appl.} {\bf 6} (1973) 217-240.
\bibitem{Kikkawa:76} K. Kikkawa, \emph{Quantum corrections in superconductor models}, \emph{Prog. Theor. Phys.} {\bf 56} (1976) 947-955.
\bibitem{Osipov:92} V. Bernard, A.\, A. Osipov, U.-G. Mei\ss ner, \emph{Consistent treatment of the bosonized Nambu-Jona-Lasinio model}, \emph{Phys. Lett. B} {\bf 285} (1992) 119-125.
\bibitem{Morais:17} J. Morais, B. Hiller, A. A. Osipov, \emph{A general framework to diagonalize vector -- scalar and axial-vector -- pseudoscalar transitions in the effective meson Lagrangian}, \emph{Phys. Lett. B} {\bf 773} (2017) 277-282.
\bibitem{Osipov:85b} M.\,K. Volkov, A.\,A. Osipov, \emph{$\pi -A_1$ transitions and quark masses in the superconductor type model}, \emph{Preprint of the Joint Institute for Nuclear Research}, Dubna P2-85-390 (1985).
\bibitem{Gasser:85} J. Gasser, H. Leutwyler, \emph{Chiral perturbation theory: Expansions in the mass of the strange quark}, \emph{Nucl. Phys. B} {\bf 250} (1985) 465-516.
\bibitem{Oakes:68} M. Gell-Mann, R.\,J. Oakes, and B. Renner, \emph{Behavior of current divergences under $SU(3)\times SU(3)$}, \emph{Phys. Rev.}  {\bf 175}, 2195-2199 (1968).
\bibitem{Weinberg:77} S. Weinberg, \emph{The problem of mass}, in A Festschrift for I.I. Rabi (Trans. New York Acad. Sci., Ser. 2, Vol. {\bf 38}, Ed. L Motz) (New York: New York Acad. of Sciences, 1977) p. 185.
\bibitem{Kaplan:86} D.\,B. Kaplan and A.\,V. Manohar, \emph{Current-mass ratios of the light quarks}, \emph{Phys. Rev. Lett.} {\bf 56} (1986) 2004-2007.
\bibitem{Leutwyler:97} H. Leutwyler, \emph{Light-quark masses}, in Masses of Fundamental Particles: Carg\' ese 1996, Ed. by M. L\' evy, J. Iliopoulos, R. Gastmans, and J.-M. G\' erard (Plenum, New York, 1997) 149-164.
\bibitem{Dashen:69} R. Dashen, \emph{Chiral $SU(3)\times SU(3)$ as a symmetry of the strong interactions}, \emph{Phys. Rev.}, {\bf 183} (1969) 1245-1260.  
\bibitem{Leutwyler:09} H. Leutwyler, \emph{Light-quark masses}, in the Proceedings of the Workshop on Chiral Dynamics, Bern, 2009, arXiv:0911.1416 [hep-ph].
\bibitem{Colangelo:18} G. Colangelo, S. Lanz, H. Leutwyler, and E. Passemar, \emph{Dispersive analysis of $\eta\to 3\pi $}, \emph{Eur. Phys. J. C} {\bf 78}, 11, 947 (2018).
\bibitem{Fodor:16} Z. Fodor et al., (Budapest-Marseille-Wuppertal Collaboration), \emph{Up and down quark masses and corrections to Dashen's theorem from lattice QCD and quenched QED}, \emph{Phys. Rev. Lett.} {\bf 117}, 8, 082001 (2016).
\bibitem{Giusti:17} D. Giusti et al., (RM123 Collaboration), \emph{Leading isospin-breaking corrections to pion, kaon and charmed-meson masses with Twisted-Mass fermions}, \emph{Phys. Rev. D} {\bf 95}, 11, 114504 (2017).
\bibitem{Duncan:96}  A. Duncan, E. Eichten, H. Thacker, \emph{Electromagnetic splittings and light quark masses in lattice QCD}, \emph{Phys. Rev. Lett.} {\bf 76} (1996) 3894-3897.
\bibitem{Leutwyler:96} H. Leutwyler, \emph{The ratios of the light quark masses}, \emph{Phys. Lett. B} {\bf 378} (1996) 313-318.
\bibitem{PDG:22} R.\,L. Workman et al. (Particle Data Group), \emph{Prog. Theor. Exp. Phys.} {\bf 2022}, 083C01 (2022).
\bibitem{Urech:95} R. Urech, \emph{$\rho^0-\omega$ mixing in chiral perturbation theory}, \emph{Phys. Lett. B} {\bf 355} (1995) 308-312.  
\bibitem{Cvetic:97} G. Cvetic, \emph{Regularization at the next-to-leading order in the top-mode standard model without gauge bosons}, Annals of Physics {\bf 255} (1997) 165-203.
\bibitem{Klimt:90} S. Klimt, M. Lutz, U. Vogl and W. Weise, \emph{Generalized $SU(3)$ Nambu -- Jona-Lasinio model (I). Mesonic modes},  Nucl. Phys. A {\bf 516} (1990) 429-468. 
\bibitem{Schifman:81} V.\,A. Novikov, M.\,A. Shifman, A.\,I. Vainshtein and V.\,I. Zakharov, \emph{Are all hadrons alike?}, Nucl. Phys. B {\bf 191} (1981) 301-369.
\bibitem{Shuryak:93} E.\,V. Shuryak, \emph{Correlation functions in the QCD vacuum}, Rev. Mod. Phys. {\bf 65} (1993) 1-46.
\bibitem{Osipov:22} A.\,A. Osipov, \emph{Light quark masses in the theory with the dynamical breaking of chiral symmetry}, \emph{JETP Letters} {\bf 115} No.6 (2021) 305-311.  
\end{thebibliography}
\end{document}